\documentclass[
floatfix,
aps,
pra,
amsmath,
twocolumn,
superscriptaddress
]{revtex4-2}

\pdfoutput=1

\usepackage{graphicx}
\usepackage{wrapfig}
\usepackage{amsmath}
\usepackage{amssymb}
\usepackage{bbm}
\usepackage{hyperref}
\usepackage{soul}
\usepackage{xcolor}

\newcommand{\ket}[1]{|{#1}\rangle}
\newcommand{\bra}[1]{\langle {#1} | }

\begin{document}

\title{Proposal for entangling gates on fluxonium qubits via a two-photon transition}

\author{Konstantin N. Nesterov}
\affiliation{Department of Physics  and Wisconsin Quantum Institute, University of Wisconsin-Madison, Madison, WI 53706, USA}

\author{Quentin Ficheux}
\affiliation{Department of Physics, Joint Quantum Institute,
and 
Center for Nanophysics and
\\ Advanced Materials,
University of Maryland, College Park, MD 20742, USA}

\author{Vladimir E. Manucharyan}
\affiliation{Department of Physics, Joint Quantum Institute,
and 
Center for Nanophysics and
\\ Advanced Materials,
University of Maryland, College Park, MD 20742, USA}

\author{Maxim G. Vavilov}
\affiliation{Department of Physics and Wisconsin Quantum Institute, University of Wisconsin-Madison, Madison, WI 53706, USA}

\date{\today}

\begin{abstract}

We propose a family of microwave-activated entangling gates on two capacitively coupled fluxonium qubits. 
A microwave pulse applied to either qubit at a frequency near the half-frequency of the  $|00\rangle - |11\rangle$ transition induces two-photon  Rabi oscillations with a negligible leakage outside the computational subspace, owing to the strong anharmonicity of fluxoniums. 
By adjusting the drive frequency, amplitude, and duration, we obtain the gate family that is locally equivalent to the fermionic-simulation gates such as $\sqrt{\rm SWAP}$-like and controlled-phase gates. 
The gate error can be tuned below $10^{-4}$ for a pulse duration under 100 ns without excessive circuit parameter matching. Given that the fluxonium coherence time can exceed 1 ms, our gate scheme is promising for large-scale quantum processors.

\end{abstract}

\pacs{}

\maketitle

\section{Introduction}

A programmable  quantum computer requires a very low error rate for the two-qubit gate operations, both for quantum error-correction schemes to work~\cite{Bravyi1998, Fowler2012}, and for extending the depth of quantum circuits during calculations on noisy intermediate-scale quantum processors~\cite{Preskill2018}. 
In the superconducting circuits platform~\cite{Devoret2013, Wendin2017, Kjaergaard2020_review}, major results were obtained using transmon qubits \cite{Blais2004,Koch2007}, which are much closer to weakly anharmonic oscillators than to two-level systems. Although simplicity and robustness of transmons facilitated the creation of processors with
dozens of qubits~\cite{Arute2019,Jurcevic2021}, the weak anharmonicity and finite coherence time have been major factors limiting gate errors. These challenges exist in both major families of two-qubit gates realized with these qubits: flux-tunable~\cite{Barends2014, Chen2014,  Collodo2020, Foxen2020, Negirneac2021, Sung2020} and microwave-activated~\cite{Leek2009, Chow2011, Chow2012, Poletto2012, Chow2013, Sheldon2016b, Krinner2020} two-qubit gates, where the gate speed is bounded by the anharmonicity. 

Fluxonium qubits~\cite{Manucharyan2009} are architecturally similar devices to transmons but they have a much stronger anharmonicity and considerably longer coherence times~\cite{Nguyen2019, Somoroff2021}. Theoretical proposals to realize microwave-activated two-qubit gates with fluxoniums and heavy fluxoniums have previously been based on driving transitions outside of the computational subspace~\cite{Nesterov2018, Abdelhafez2020}. 
Recent experiments demonstrated  fast two-qubit gates on fluxoniums activated by driving close to such transitions~\cite{Ficheux2021, Xiong2021}.  Because these noncomputational transitions generally have  shorter lifetimes than the computational ones, such a gate scheme is exposed to additional error channels. A flux-tunable entangling gate with fluxonium qubits has also been recently reported~\cite{Wang2021a}, but it is subject to extra dephasing errors when a qubit is moved away from its flux sweet spot. 

In this work we consider a gate that keeps the state entirely in the computational subspace with qubits parked at their sweet spots, hence benefiting in full from the long coherence of fluxonium qubits. 
The entangling gate presented here is accomplished by a high-power microwave drive at half the frequency of the $\ket{00} - \ket{11}$ transition, which induces two-photon transitions between $\ket{00}$ and $\ket{11}$ and activates a coherent  mixing in the $\{\ket{00}, \ket{11}\}$ subspace.
With  weak interaction in the computational subspace, the gate would normally be slow. However, the strong anharmonicity of fluxoniums makes it possible to perform fast gate operations by increasing the drive amplitude without generating leakage to noncomputational levels. For realistic fluxonium parameters, we demonstrate that a 50-ns-long gate with $\pi/2$ mixing angle in the $\{\ket{00}, \ket{11}\}$ subspace can be realized with the leakage error below $10^{-4}$ and total gate error below $10^{-3}$ without using advanced pulse shaping for presently achievable coherence times~\cite{Nguyen2019, Somoroff2021}. For longer gates, about 100 ns long, the coherent gate error can be reduced below $10^{-4}$. With decoherence effects accounted for, the $10^{-4}$ threshold requires some improvement of the best existing lifetimes~\cite{Nguyen2019, Somoroff2021} and should be possible in next-generation devices.

The entangling power \cite{Zanardi2000, Ma2007}  and  the local equivalence class of the proposed gates depend on the mixing angle in  the $\{\ket{00}, \ket{11}\}$ subspace and on the magnitude of the effective $ZZ$ coupling. The $ZZ$ coupling originates from both the static level repulsion and an induced ac-Stark shift due to the large drive amplitude.  This term  modifies entangling properties of the gate.
Interestingly, we show that for a half rotation in the $\{\ket{00}, \ket{11}\}$ subspace, the entangling power of the gate is independent of the  contribution due to the total $ZZ$ coupling. In fact, the family of gates with a half-rotation contains gates  locally equivalent to $\sqrt{\rm iSWAP}$ and $\sqrt{\rm SWAP}$. 

Entangling gates activated by two-photon processes were  proposed and implemented in trapped-ion systems~\cite{Sorensen2000, Benhelm2008}. 
In superconducting systems, a two-photon gate based on driving the $\ket{00} - \ket{11}$ transition was demonstrated  experimentally with transmon qubits~\cite{Poletto2012}. 
This gate  required frequency matching between $\ket{0}-\ket{1}$ transition of one transmon and $\ket{1}-\ket{2}$ transition of another transmon to increase the two-photon Rabi frequency by increasing the hybridization of the $\ket{11}$ state with one of the noncomputational states. In the case of fluxoniums, we  can speed up the two-photon Rabi oscillations by increasing the drive amplitude. This trick does not cause leakage outside the computational subspace because of the strong anharmonicity of fluxoniums.
 Thus, the scheme presented here does not require frequency matching. 
In general, our gate benefits from higher frequencies of qubit transitions, which lead to a stronger hybridization of states $|01\rangle$ and $|10\rangle$ and, therefore, to a higher two-photon Rabi frequency. From practical considerations, the suggested qubit frequency range is around $0.5 - 1~\textrm{GHz}$.

While a single entangling gate combined with individual qubit controls is sufficient to generate a universal set of logical operations, some algorithms may compile more efficiently with a larger two-qubit gate set, especially if this set is hardware efficient~\cite{Kivlichan2018, Lacroix2020}. For a given algorithm executed on a noisy processor, the maximal depth of a quantum circuit depends on a particular set of gates implemented on the hardware level. The gates based on mixing of $\ket{00}$ and $\ket{11}$ discussed here are locally equivalent to and can be easily mapped by single-qubit  $X$ rotations into  the operations in the $\{\ket{01}, \ket{10}\}$ subspace. 
The family of the proposed gates 
is equivalent to  a complete set of fSim gates for the fermionic-simulation problem~\cite{Abrams2019, Foxen2020}.

The outline of the paper is as follows. In Sec.~\ref{Sec:model}, we introduce our model  and discuss relevant spectral properties of fluxonium circuits. In Sec.~\ref{Sec:two-photon-trans}, we consider coherent two-photon transitions between $\ket{00}$ and $\ket{11}$ two-qubit states and calculate their rate both analytically and numerically. In Sec.~\ref{Sec:gates}, we analyze two-qubit gates realized via the two-photon $\ket{00}-\ket{11}$ transition. We discuss local equivalence classes of such gates and their entangling power and simulate the gate error. We conclude in Sec.~\ref{Sec:conclusions}.
 
\section{Capacitively coupled fluxonium qubits}\label{Sec:model}

\begin{figure}[t]
\includegraphics[width=\columnwidth]{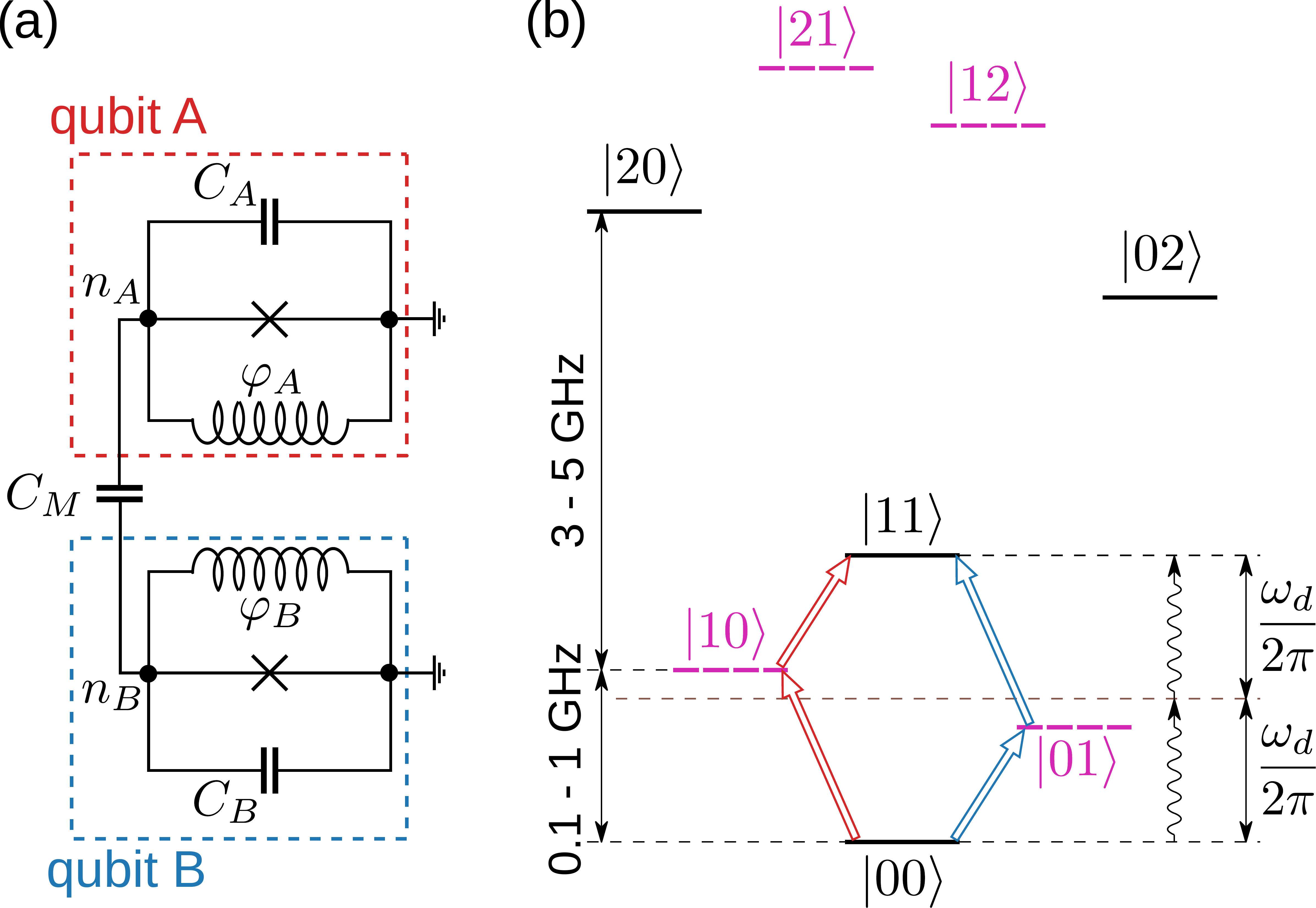}\caption{
(a) Circuit diagram of two capacitively coupled fluxonium qubits. (b) Energy levels (thick horizontal lines) of the coupled system are separated into two groups based on whether or not single-qubit indices have the same parity  (black solid and magenta dashed lines). Transitions between different parity groups can be performed with a single microwave photon, while transitions within the same group require higher-order multiphoton processes.   In the perturbative regime, the coherent two-photon transition between $|{00}\rangle$ and $|{11}\rangle$ is dominated by contributions from two virtual states generated by states $|{10}\rangle$ and $|{01}\rangle$, as shown by the wide arrows. These contributions interfere destructively and exactly cancel each other  when the coupling vanishes. 
}\label{Fig-explanation}
\end{figure}

The circuit diagram of two coupled fluxoniums, labeled as $A$ and $B$, is shown schematically in Fig.~\ref{Fig-explanation}(a).  We model this system by the Hamiltonian
\begin{equation}\label{Hamiltonian-two-qubit}
\hat{H} = \hat{H}^{(0)}_{A} + \hat{H}^{(0)}_B + \hat{V} + \hat{H}_{\rm drive}\,,
\end{equation}
where
\begin{equation}\label{Hamiltonian-fluxonium}
 \hat{H}_{\alpha}^{(0)} = 4E_{C,\alpha} \hat{n}_\alpha^2 + \frac 12 E_{L,\alpha} \hat{\varphi}_\alpha^2 - E_{J,\alpha} \cos(\hat{\varphi}_\alpha - \phi_{\rm ext,\alpha})\,
\end{equation}
describes individual qubits ($\alpha = A, B$)~\cite{Manucharyan2009}. The capacitive interaction between qubits is given by 
\begin{equation}\label{interaction-charge}
 \hat{V} = J_C \hat{n}_A \hat{n}_B \,,
\end{equation}
and the coupling to an external microwave drive of frequency $\omega_d$  and phase  $\gamma_d$ -- by
\begin{equation}\label{drive}
 \hat{H}_{\rm drive} =2\hbar f(t)\cos(\omega_d t + \gamma_d) \left(\eta_A\hat{n}_A +  \eta_B\hat{n}_B\right) \,.
\end{equation}
Here $\hbar = h/2\pi$ is the reduced Planck constant, $f(t)$ is the time-dependent field envelope, and $\eta_A$ and $\eta_B$ describe the coupling of each qubit to the driving field.

In these equations, the canonical variables are the dimensionless flux $\hat{\varphi}_\alpha$ and charge (the number of Cooper pairs) $\hat{n}_\alpha$, which satisfy the commutation relations $[\hat{\varphi}_\alpha, \hat{n}_{\alpha'}] = i\delta_{\alpha\alpha'}$. The kinetic-energy term in Eq.~(\ref{Hamiltonian-fluxonium}) is determined by the charging energy $E_{C, \alpha} = e^2/2C_\alpha$, where $(-e)$ is the electron charge and $C_\alpha$ is the total capacitance of the circuit $\alpha$. The  inductive energy is $E_{L, \alpha} = (\hbar / 2e)^2 / L_\alpha$, where $L_\alpha$ is the effective linear inductance of a long chain of Josephson junctions, which is a hallmark of the fluxonium.  This superinductance is shunted by a small  junction, associated with the Josephson energy $E_{J, \alpha}$. 
The final term in Eq.~(\ref{Hamiltonian-fluxonium})  depends on $\phi_{{\rm ext}, \alpha} = (2e/\hbar)\Phi_{\rm ext, \alpha}$, where  $\Phi_{\rm ext, \alpha}$ is the magnetic flux threading the loop formed by the small junction and superinductance. In the limit of  $C_M \ll C_A, C_B$, where $C_M$ is the mutual capacitance, the interaction strength in Eq.~(\ref{interaction-charge}) is given by
$J_C = 4e^2{C_M}/\left({C_A C_B}\right)$~\cite{Vool2017, Nesterov2018}.

Below, we assume that fluxonium circuits are at their half-flux-quantum sweet spots defined by $\phi_{\rm ext,\alpha} = \pi$, where the circuits are first-order insensitive to the external flux noise~\cite{Manucharyan2012}. We label eigenstates of Hamiltonian~(\ref{Hamiltonian-fluxonium}) as $|0_\alpha\rangle, |1_\alpha\rangle, |2_\alpha\rangle, \ldots$ in  increasing order of the corresponding eigenenergies $E^\alpha_0 \leq E^\alpha_1 \leq \ldots$.  The first two levels of each circuit $\alpha$ form a qubit with  transition frequency $\omega^\alpha_{01}$, where we define single-fluxonium frequencies as $\hbar\omega^\alpha_{kl} = E^\alpha_{l} - E^\alpha_k$.
The qubit transition $|0_\alpha\rangle  - |1_\alpha\rangle$ can display an exceptionally long coherence time exceeding 500 $\mu$s~\cite{Nguyen2019}, which makes it an attractive choice for quantum-information storage. 
 The qubit transition frequency $\omega^\alpha_{01}/2\pi$ is typically in the 100 MHz - 1 GHz range, which is much lower than conventional values in superconducting qubits. In addition, the charge matrix element $n^\alpha_{01}$ with the notation $n^\alpha_{kl} = |\langle k_\alpha|\hat{n}_\alpha| l_\alpha\rangle|$ is suppressed at low frequencies. At the same time, the transition $|1_\alpha\rangle  - |2_\alpha\rangle$ has properties similar to those of the transmon  with a typical frequency of several gigahertz~\cite{Manucharyan2012, Zhu2013b, Lin2018}. Because of the potential-energy symmetry at $\phi_{\rm ext,\alpha} = \pi$, the matrix elements of $\hat{n}_\alpha$ display parity selection rules, e.g., $n^\alpha_{02}=n^\alpha_{13}=0$~\cite{Zhu2013b, Zhu2013a, Nesterov2018}. However, $n^\alpha_{03}$ is not suppressed and can be of the order of $n^\alpha_{12}$~\cite{Zhu2013a, Zhu2013b}.

We label interacting (dressed) two-qubit eigenstates of the Hamiltonian~(\ref{Hamiltonian-two-qubit}) at $\hat{H}_{\rm drive}=0$ as $|{kl}\rangle$ implying adiabatic connection to the  noninteracting tensor-product  states $|kl\rangle_0 = |k_A\rangle  |l_B\rangle $. The frequency of the two-qubit transition $|{kl}\rangle  - |{k'l'}\rangle$ is denoted as $\hbar\omega_{kl  - k'l'} = E_{k'l'} - E_{kl}$, where $E_{kl}$ is the eigenenergy of $|{kl}\rangle$. The two-qubit computational subspace $\{|{00}\rangle, |{01}\rangle,
|{10}\rangle, |{11}\rangle\}$ is well separated from higher levels as illustrated schematically in Fig.~\ref{Fig-explanation}(b). 
 Two-qubit levels  can be divided into two parity groups depending on whether $k+l$ is even or odd, which is shown with solid and dashed lines in Fig.~\ref{Fig-explanation}(b). Because of the parity selection rules for the charge operators $\hat{n}_A$ and $\hat{n}_B$, the interaction term (\ref{interaction-charge}) mixes levels within the same parity group only, while the matrix elements of the drive term (\ref{drive}) are nonzero only between levels belonging to different parity groups~\cite{Nesterov2018}.

\section{Two-photon Rabi oscillations}\label{Sec:two-photon-trans}

In this section, we consider a continuous drive of the two fluxoniums with a constant amplitude $f$ in Eq.~\eqref{drive} and a drive frequency about half the frequency of the $\ket{00}$ -- $\ket{11}$ transition. Understanding this process is essential for the two-qubit gate discussed in Sec.~\ref{Sec:gates}.

\subsection{Rabi frequency}\label{Sec:perturbative-analysis}

Even though two fluxonium excitations cannot be created by a single microwave photon, i.e., $\langle {00}|\hat{H}_{\rm drive}|{11}\rangle=0$ because of the parity selection rules, the microwave drive can still induce the $|{00}\rangle - |{11}\rangle$ transition  exchanging two fluxonium excitations with pairs of drive photons.
The transition amplitude calculated to the leading (second) order in $f$ can be understood as having contributions from  cascaded sequential single-photon transitions via intermediate real states such as $\ket{01}$ and $\ket{10}$ and from coherent two-photon processes via intermediate virtual states~\cite{Foot2005_book}. 
Under certain conditions discussed below, the excitation probabilities of states $\ket{01}$ and $\ket{10}$ can remain low, while the system state oscillates between $\ket{00}$ and $\ket{11}$ with high visibility. Below, we apply a perturbation theory to estimate these probabilities and the frequency of the two-photon Rabi oscillations between $\ket{00}$ and $\ket{11}$.
We  approximate fluxoniums as two-level systems, which is reasonable given their strong anharmonicity. Nonperturbative effects and higher fluxonium levels  are accounted for in numerical analysis of Sec.~\ref{Sec:Rabi-numerics}.

The resonant Rabi frequency for a single-photon transition $|kl\rangle  - |k'l'\rangle$ such as $|00\rangle  - |10\rangle$ is  given by the matrix elements of the drive [see Eq.~\eqref{drive}]:
\begin{subequations}
\begin{equation}\label{Omega-one-photon}
\Omega_{kl  - k'l'} = 2 f |N_{kl, k'l'}|\,,
\end{equation}
where
\begin{equation}
\label{Nkl}
N_{kl, k'l'}  = \langle {kl}|(\eta_A\hat{n}_A +  \eta_B\hat{n}_B)| {k'l'}\rangle\,.
\end{equation} 
\end{subequations}
Without qubit interactions, the eigenstates  of the system are product states of each qubit, $\ket{kl}_0 = \ket{k_A}\otimes\ket{l_B}$, and the Rabi frequencies \eqref{Omega-one-photon} reduce to the single-qubit Rabi frequencies for the $\ket{0_\alpha}-\ket{1_\alpha}$ transitions 
\begin{equation}
\Omega_{\alpha,0} = 2f \eta_\alpha \left| \bra{1_\alpha} \hat{n}_\alpha\ket{0_\alpha}\right|\quad
\{\alpha= A,B\}\,.
\end{equation}
Here we consider microwave drives with frequencies close to
\begin{equation}
\label{baromega}
\bar\omega = \frac{\omega_{00-11}}{2}\,,
\end{equation}
which reduces to the average of qubit frequencies $(\omega_{01}^A+\omega_{01}^B)/{2}$ when $J_C=0$.
When driving with frequency $\omega_d = \bar \omega$, 
the probability of single-fluxonium transitions is bounded from above by the contrast of Rabi oscillations in two-level systems. The latter is given by
\begin{equation}\label{contrast-single-photon}
P_{\alpha} = 
\frac{\Omega_{\alpha,0}^2}{\Omega_{\alpha,0}^2 + (\Delta_{AB}/2)^2}\,,
\end{equation}
where $\Delta_{AB}= |\omega^A_{01} - \omega^B_{01}|$ is the detuning between qubit frequencies.
Taking
$P_{\alpha} \ll 1$, we obtain conditions on the drive amplitudes,
\begin{equation}\label{Omega_alpha_ll_Delta}
\Omega_{\alpha,0} \ll \Delta_{AB}\,,
\end{equation}
which combined with $\omega_d= \bar\omega$ imply that
correlated two-qubit oscillations will dominate the dynamics of the system over independent single-qubit excitations.

We now consider the transition between $\ket{00}$ and $\ket{11}$, which is activated by two-photon processes when the drive frequency is $\bar{\omega}$.
In this case, the system exhibits full oscillations of its probability being in one of the states, and the frequency $\widetilde{\Omega}$ of such two-photon Rabi oscillations depends on the matrix element of the two-photon drive between $\ket{00}$ and $\ket{11}$.
We apply second-order  perturbation theory together with the rotating-wave approximation (RWA) to obtain
\begin{equation}\label{Omega-0011-pert}
 \widetilde{\Omega}
= \frac{\left|
\Omega_{00-01} \Omega_{01- 11}  - \Omega_{00- 10} \Omega_{10- 11}
\right|}{\widetilde{\Delta}_{AB}}
\,,
\end{equation}
where 
\begin{equation}
    \widetilde{\Delta}_{AB} = |\omega_{00-01} - \omega_{00-10}|\,,
\end{equation}
which differs from $\Delta_{AB}$ by a correction quadratic in $J_C$. 
This equation is reminiscent of a similar result derived for trapped ions excited with a bichromatic laser~\cite{Sorensen2000, Benhelm2008}.
By choosing $\omega_d=\bar\omega$ in the derivation of Eq.~\eqref{Omega-0011-pert}, we 
neglected the shift of the qubit frequencies due to the Stark shifts, which are quadratic in the drive amplitude $f$. 
Equation~(\ref{Omega-0011-pert}) describes destructive interference between the two contributions corresponding to two paths via virtual states generated by  states $|{01}\rangle$ and $|{10}\rangle$,  indicated by arrows in Fig.~\ref{Fig-explanation}(b). Without interaction, $J_C=0$, we observe that Rabi frequencies $\Omega_{kl- k'l'}$ for single-photon transitions reduce to single-qubit Rabi frequencies $\Omega_{A, 0}$ and $\Omega_{B, 0}$,
and Eq.~\eqref{Omega-0011-pert} yields $\widetilde{\Omega}=0$. This result emphasizes that   entanglement is impossible without qubit interactions.

\subsection{Interaction effects}

Let us now calculate the two-photon Rabi frequency to the first nonvanishing order in $J_{\rm eff} = J_C |n^A_{01} n^B_{01}| \ll \hbar\Delta_{AB}$. Even though $J_C$ can be large,  $J_{\rm eff}\ll J_C$ because $|n^A_{01} n^B_{01}|$ is typically small for fluxonium qubits. 
Correction to the denominator of Eq.~\eqref{Omega-0011-pert} is quadratic in $J_C$, while corrections to matrix elements are linear. Thus, we find that $\widetilde{\Omega}$ is finite because of the hybridization of $|01\rangle$ with $|10\rangle$ and of $|00\rangle$ with $|11\rangle$. The interaction-dressed eigenstates in the first pair of states have the form
\begin{equation}\label{states01_10_J_perturbative}
|{01}\rangle = |01\rangle_0 - \frac{J_{\rm eff}}{\hbar\Delta_{AB}}|10\rangle_0\,, \quad
|{10}\rangle = |10\rangle_0 + \frac{J_{\rm eff}}{\hbar\Delta_{AB}}|01\rangle_0\,.
\end{equation}
The mixing amplitude for the pair  of states $\ket{00}$ and $\ket{11}$ 
has a form similar to Eq.~\eqref{states01_10_J_perturbative}, but with $2\bar\omega$ in the denominator instead of $\Delta_{AB}$.  Thus, the mixing of states $\ket{00}$ and $\ket{11}$ by interaction is 
reduced by  $\Delta_{AB} / 2\bar\omega$ compared to that of states $\ket{01}$ and $\ket{10}$. This factor is 
not necessarily small in fluxonium qubits. Nevertheless, we ignore it for now to focus on  the main principles.  We  find  the following expression for the two-photon Rabi frequency:
\begin{equation}\label{Omega-0011-linear-JC}
\widetilde \Omega
= 2  n^A_{01} n^B_{01} \frac{J_C}{\hbar}
\frac{\Omega_{A,0}^2 +\Omega_{B,0}^2}{\Delta^2_{AB}}\,.
\end{equation}

The two-photon rate increases with   hybridization in the computational subspace. One natural way to increase the latter is to reduce the detuning $\Delta_{AB}$, which, however, enhances the magnitude of spurious single-photon excitations when driving at $\omega_d \approx \bar\omega$; see Eq.~\eqref{contrast-single-photon}. To ensure the predominance of the two-photon process over single-photon excitations, we fix the dimensionless amplitude $\Omega_{\alpha,0}/\Delta_{AB}$ rather than  $f$ since the dimensionless amplitude fully determines the relative importance of single-photon excitations, see Eq.~(\ref{contrast-single-photon}). We conclude that reducing $\Delta_{AB}$ is not practical for increasing the rate of the two-photon $|{00}\rangle - |{11}\rangle$ transition. Equation ~(\ref{Omega-0011-linear-JC}) suggests that it is beneficial to have larger values of single-qubit charge matrix elements $n^\alpha_{01}$. Because  $n^\alpha_{01} = \omega^\alpha_{01} |\langle 0|\hat{\varphi}_\alpha|1\rangle_\alpha| / 8E_{C, \alpha}$~\cite{Nesterov2018}, this condition  means that having higher qubit frequencies is beneficial to make the two-photon Rabi oscillations faster.

In addition to inducing coherent two-photon oscillations, a strong drive at $\omega_d = \bar\omega$ induces $ZZ$ interactions via the ac-Stark effect (static $ZZ$ interaction is absent in two-level models~\cite{Zhao2020, Ku2020}).
In particular, we evaluate  the relative phase accumulated in the computational subspace for a constant drive during time $t$:
\begin{equation}
    \zeta =\frac{\Delta E_{01} + \Delta E_{10} - \Delta E_{00} - \Delta E_{11}}{\hbar}  t \,,
\end{equation}
where $\Delta E_{kl}$ is the energy shift of level $|{kl}\rangle$ due to the ac-Stark effect. 
Using  second-order perturbation theory for the energy shifts
due to the  drive
combined with first-order corrections in the coupling rate,   Eq.~(\ref{states01_10_J_perturbative}), we find that
\begin{equation}
\begin{split}
\zeta & = \frac{\Omega_{00-10}^2 + \Omega_{10- 11}^2 - \Omega_{00-01}^2 
- \Omega_{01-11}^2}{\Delta_{AB}} t \\  
& =  
\frac{4\Omega_{A,0}\Omega_{B,0}}{\Omega_{A,0}^2 +\Omega_{B,0}^2}
\widetilde{\Omega} t 
\,.
\label{eq:alpha}
\end{split}
\end{equation}
In general, this phase accumulation is not negligible during a full Rabi period $t=2\pi/\widetilde \Omega$.  Note that Eq.~\eqref{eq:alpha}, obtained to first order in interaction $J_c$, vanishes if the drive is applied to only one qubit, so either $\eta_A = 0$ or $\eta_B=0$ in Eq.~\eqref{drive}.

A more rigorous analytic treatment of the two-photon process for fluxonium qubits is unnecessarily cumbersome.  Below, we present a detailed numerical analysis of the two-photon Rabi oscillations in a system of two fluxoniums.

\subsection{Numerical simulations of Rabi oscillations}\label{Sec:Rabi-numerics}
\begin{table*}[t]
    \centering
    \begin{tabular}{|c|c|c|c|c|c|c|c|c|c|}
    \hline
         & $E_{L,\alpha}/h$ & $E_{C,\alpha}/h$ & $E_{J,\alpha}/h$ & $\omega^\alpha_{01}/2\pi$ & $\omega^\alpha_{12}/2\pi$
         & $\omega^\alpha_{23}/2\pi$ 
         & $|\langle 0_\alpha |\hat{n}_\alpha |1_\alpha\rangle |$ & $|\langle 1_\alpha |\hat{n}_\alpha |2_\alpha\rangle |$
         & $|\langle 0_\alpha |\hat{n}_\alpha |3_\alpha\rangle |$
         \\
         \hline
         qubit A & 1.5 GHz & 1.0 GHz & 3.8 GHz & 1.152 GHz & 3.280 GHz 
         & 3.253 GHz & 0.249 & 0.608 & 0.260 
         \\
         qubit B & 0.9 GHz & 1.0 GHz & 3.0 GHz & 0.849 GHz & 2.929 GHz 
         & 2.683 GHz & 0.207 & 0.567 & 0.277
         \\
         \hline
    \end{tabular}
    \caption{Fluxonium parameters used for numerical simulations.}
    \label{Table-params}
\end{table*}

Based on Eq.~\eqref{Omega-0011-linear-JC}, we choose the  single-qubit parameters for all numerical simulations in this paper as shown in Table~\ref{Table-params}. 
With these values, both main qubit transition frequencies are relatively high compared to usual fluxonium qubits, which is accompanied by larger values of 0-1 charge matrix elements.
In simulations, we first diagonalize Hamiltonians of single fluxonium circuits~(\ref{Hamiltonian-fluxonium}), and then work with the interacting system taking five lowest levels in each fluxonium. We use the full Hamiltonian~\eqref{Hamiltonian-two-qubit} in the laboratory frame and, therefore, go beyond the RWA in addition to going beyond the perturbation theory and two-level approximation used in Eq.~\eqref{Omega-0011-pert}.  For simplicity, we only consider the case $\eta_A = \eta_B = 1$ here. 

\begin{figure}[t]
\includegraphics[width=\columnwidth]{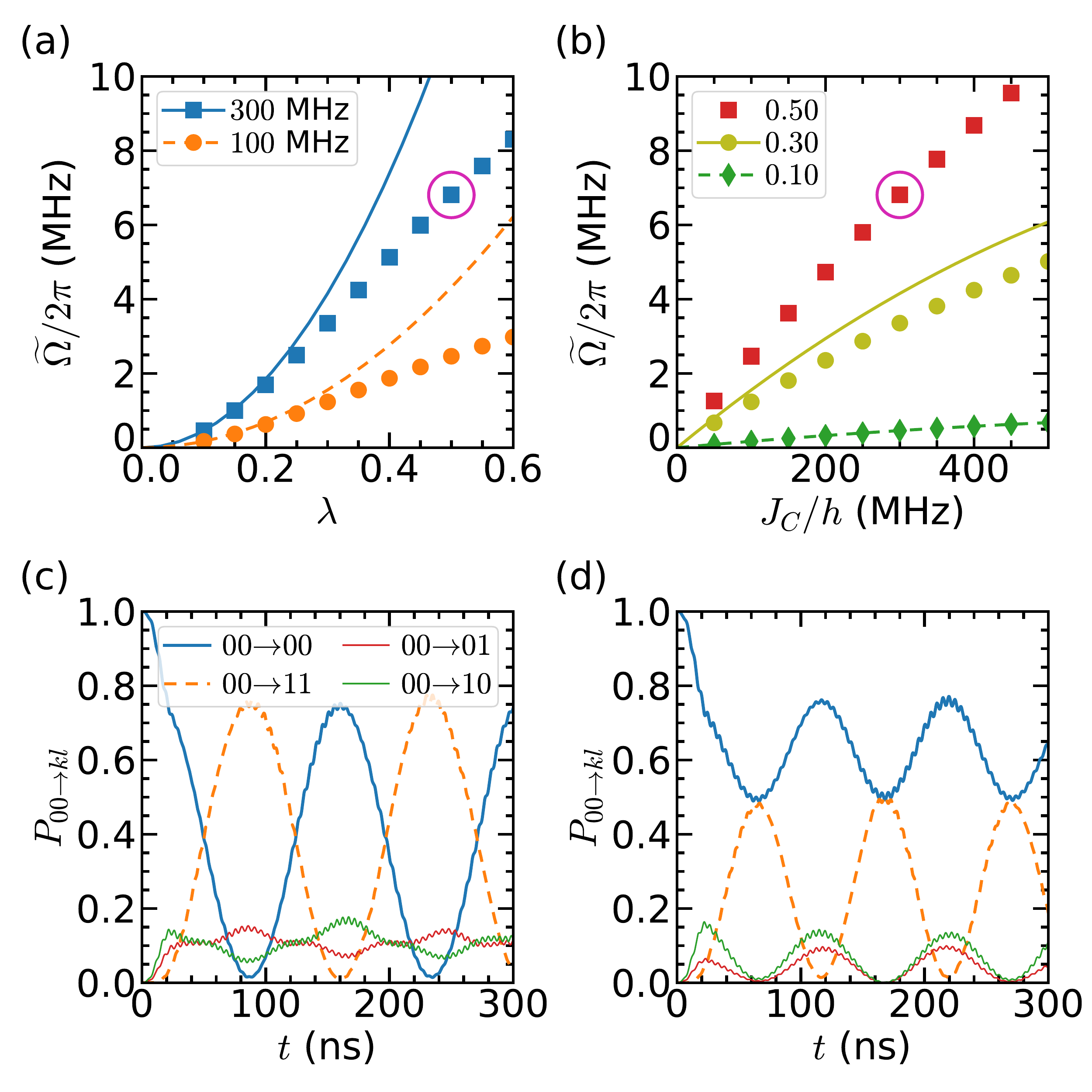}\caption{
(a), (b) The Rabi frequency of the two-photon $|{00}\rangle - |{11}\rangle$ transition vs the dimensionless drive amplitude $\lambda$ [see Eq.~(\ref{lambda})] at $J_C/h=100$ and $300$ MHz (a) and vs  $J_C/h$ at $\lambda=0.1$, $0.3$, and $0.5$ (b). Lines show perturbative calculations using Eq.~(\ref{Omega-0011-pert}), and symbols show numerical simulations with the drive frequency $\omega_d$ chosen to maximize the contrast of Rabi oscillations. (c) Time evolution of computational-basis populations with  drive parameters corresponding to the circled data point in panel (a) and (b); the starting state is $|{00}\rangle$. (d) Same as in panel (c), but with a different $\omega_d$, 
which yields oscillations that are approximately $\sqrt{2}$ faster than in panel (c).}
\label{Fig-rabi-osc}
\end{figure}

We plot the Rabi frequency for the $|{00}\rangle - |{11}\rangle$ transition as a function of the dimensionless drive amplitude 
\begin{equation}
\label{lambda}
\lambda = \frac{\Omega_{A,0}}{\Delta_{AB}},    
\end{equation}
see Fig.~\ref{Fig-rabi-osc}(a), and interaction strength $J_C/h$, see Fig.~\ref{Fig-rabi-osc}(b). Lines show calculations based on the analytic result~(\ref{Omega-0011-pert}) except that multilevel fluxonium qubits were used to compute energies and matrix elements, and symbols show results of numerical simulations, which were performed as follows. For given $\lambda$ and $J_C$, we chose $\omega_d$ that maximizes the contrast between the minimum of $P_{00 \to 00}(t)$ and maximum of $P_{00  \to 11}(t)$, where $P_{kl \to k'l'}(t)$ is the population of state $|{k'l'}\rangle$  at time $t$ for the initial state $|{kl}\rangle$ at $t=0$. At each $\omega_d$, we calculated these probabilities via time-domain simulations at time $t\ge 0$. In the drive term~(\ref{drive}), we used a pulse with a Gaussian rising edge at $0<t<t_{\rm rise} = 25$ ns, which is given by
\begin{equation}\label{pulse-shape}
    f(t) \propto \exp\left[-\frac{(t - t_{\rm rise})^2}{2\sigma^2}\right] - \exp\left[-\frac{t_{\rm rise}^2}{2\sigma^2}\right]\,,
\end{equation}
where $\sigma = t_{\rm rise}/2$, and with the amplitude of the flat part ($t>t_{\rm rise}$) determined by $\lambda$ via Eq.~\eqref{lambda}. An example of such time-domain simulations of the occupation probabilities for $\omega_d$ that maximizes the contrast is shown in Fig.~\ref{Fig-rabi-osc}(c) for $\lambda=0.5$ and $J_C/h=300$ MHz. The optimal contrast in Fig.~\ref{Fig-rabi-osc}(c) is approximately 80\%,  which is different from 100\% because of a finite $t_{\rm rise}$ in Eq.~\eqref{pulse-shape} and because of leaking single-photon transitions such as $\ket{00}- \ket{01}$. This contrast is at least 75\% for all the pairs of $\lambda$ and $J_C$ discussed in Fig.~\ref{Fig-rabi-osc} and is close to 100\% for $\lambda \lesssim 0.1$. The observed period of oscillations in Fig.~\ref{Fig-rabi-osc}(c) is 147 ns, which corresponds to a Rabi frequency of 6.8 MHz in agreement with the circles in Figs.~\ref{Fig-rabi-osc}(a) and \ref{Fig-rabi-osc}(b).  In a two-level system with qubit A parameters, $\lambda=0.5$  corresponds to the 50\% contrast of single-photon Rabi oscillations, see Eq.~(\ref{contrast-single-photon}), which agrees with appreciable occupations of states $|{10}\rangle$ and $|{01}\rangle$ in Fig.~\ref{Fig-rabi-osc}(c).

Both Figs.~\ref{Fig-rabi-osc}(a) and \ref{Fig-rabi-osc}(b) demonstrate agreement between analytic and numerical calculations  at $\lambda \lesssim 0.25$. On the one hand, this agreement is not surprising since Eq.~\eqref{Omega-0011-pert} is based on the perturbation theory and RWA in the computational subspace, which are both valid at small $\lambda$. On the other hand,  Eq.~(\ref{Omega-0011-pert}) was derived for two-level systems while Fig.~\ref{Fig-rabi-osc} presents results for the full fluxonium Hamiltonian~\eqref{Hamiltonian-fluxonium}. 
To elaborate on the effect of higher levels, we introduce an analog of the dimensionless drive amplitude $\lambda$ for the $\ket{1_A}-\ket{2_A}$ transition of qubit A with similar reasoning applied to other transitions:
\begin{equation}\label{lambda_1_2}
    \lambda_{1-2} = \frac{\Omega_{A, 0}^{1-2}}{2(\omega^{A}_{12} - \bar{\omega})}  \,,
\end{equation}
where $\Omega_{A, 0}^{1-2} = 2f n^A_{12} $ is the corresponding single-qubit Rabi frequency.
For the parameters of Table~\ref{Table-params},  $\lambda_{1-2}\approx 0.16 \lambda$, implying that the single-photon transition $\ket{1_A}-\ket{2_A}$ is suppressed for all the values of $\lambda$ discussed in Fig.~\ref{Fig-rabi-osc}(a). For generic fluxonium parameters, we  do not expect $\lambda_{1-2}/\lambda$ to become large. Therefore, even with higher levels taken into account, $\lambda \ll 1$ ensures that  the $\ket{00}-\ket{11}$ transition is still dominated by coherent two-photon processes via intermediate virtual states with suppressed sequential one-photon transitions via real states. However, a higher-energy noncomputational level $|{kl}\rangle$ ($k>1$ or $l>1$) formally generates an additional virtual state, producing an extra term in Eq.~(\ref{Omega-0011-pert}) as long as $N_{00, kl}N_{kl, 11}\ne 0$, where $N_{kl, k'l'}$ is defined in Eq.~\eqref{Nkl}. Such terms are exactly zero at $J_C=0$ and acquire finite values at $J_C\neq 0$. For the parameters of Table~\ref{Table-params}, they are negligible in comparison to existing contributions because their denominators, which are determined by $\omega_{00 - kl}-\bar \omega$, are much larger than $\Delta_{AB}$. This is not the case in general because possibly large values of $n^\alpha_{12}$ or $n^\alpha_{03}$ may make 
$N_{00, kl}N_{kl, 11}$ to be sufficiently large and the corresponding additional contribution to Eq.~(\ref{Omega-0011-pert}) non-negligible in comparison to the sum of existing terms, which interfere destructively. Equation~(\ref{Omega-0011-pert}) works for our choice of parameters when the higher-energy states do not contribute significantly, but this approximation may be less accurate for other choices of fluxonium parameters, e.g., for qubits with larger $n^\alpha_{12}/n^\alpha_{01}$ or $n^\alpha_{03}/n^\alpha_{01}$.

At $\lambda \gtrsim 0.25$, perturbation theory breaks down and the Rabi frequency increases slower than  $\lambda^2$. We emphasize that $\widetilde{\Omega}$ still increases monotonically even far from the perturbative regime, when single-photon transitions can be strongly excited. It can become as large as a few megahertz even for a relatively small $J_C/h$ of $100$ MHz and can surpass 10 MHz for stronger interaction strengths. We note that  $\widetilde \Omega$ is close to being linear in $J_C$, which qualitatively agrees with Eq.~(\ref{Omega-0011-linear-JC}).   Accounting for hybridization between computational and higher-energy  states is necessary for a quantitative agreement.

The ability to induce two-photon Rabi oscillations can be used to realize an entangling gate involving the  mixing of states $|{00}\rangle$ and $|{11}\rangle$ similar to bSWAP of Ref.~\cite{Poletto2012}. 
A proper pulse shape is required to minimize single-photon transitions [e.g., $|{00}\rangle  - |{10}\rangle$ in Fig.~\ref{Fig-rabi-osc}(c)] at the end of the pulse. 
In addition, driving the $\ket{00}-\ket{11}$ transition with a different $\omega_d$ creates off-resonant Rabi oscillations that exchange only a fraction of the population between states $\ket{00}$ and $\ket{11}$. In Fig.~\ref{Fig-rabi-osc}(d), $\omega_d$ is chosen to create oscillations of $\simeq 50\%$ of the state populations. Here, the minima of $P_{00 \to 01}(t)$ and $P_{00 \to 10}(t)$ occur at times where the state of the system is in an equal superposition of $\ket{00}$ and $\ket{11}$. This feature is not a generic property of our gate, but it results from our choice of values for the parameters $\lambda$ and $J_C$.
The resulting period of oscillations is 103 ns, which is approximately $\sqrt{2}$ shorter than 147 ns in Fig.~\ref{Fig-rabi-osc}(c). This behavior is reminiscent of that of  a driven two-level system; the period of detuned Rabi oscillations with 50\% contrast is exactly $\sqrt{2}$ times shorter than the period of resonant Rabi oscillations at a fixed drive amplitude.  It  supports our understanding of the high-power drive of $|{00}\rangle  - |{11}\rangle$ at $\bar\omega$ in a coupled-fluxonium system in terms of two-photon Rabi oscillations.  Note, however, that in a true two-level system, 50\% contrast of Rabi oscillations requires a frequency detuning equal to the resonance Rabi frequency. While this Rabi frequency is 6.8 MHz for Fig.~\ref{Fig-rabi-osc}(c), the difference between the values of $\omega_d$ in Figs.~\ref{Fig-rabi-osc}(c) and \ref{Fig-rabi-osc}(d) is $4.5$ MHz. Therefore, our reasoning in terms of Rabi oscillations in a driven two-level system is correct only qualitatively, while an accurate description of the dynamical behavior requires accounting for  other levels. 

\section{Entangling gates}\label{Sec:gates}

\subsection{Theoretical concepts}\label{Sec:gates-concepts}

We parameterize the family of gates spanned by the coherent mixing of $|{00}\rangle$ and $|{11}\rangle$ states and controlled-phase operations by two angles as
\begin{equation}\label{U_gate_general}
    \hat{U}_{00-11}(\theta, \zeta) = 
    \begin{pmatrix}
    \cos\frac{\theta}{2} & 0 & 0 & -{i}\sin\frac{\theta}{2} \\
    0 & e^{i\zeta/2} & 0 & 0 \\
    0 & 0 & e^{i\zeta/2} & 0 \\
    -i \sin\frac{\theta}{2} & 0 & 0 & \cos\frac{\theta}{2}
    \end{pmatrix}\,.
\end{equation}
In the absence of leakage outside of the computational subspace and provided that single-photon processes are negligible, any two-photon process described in the previous section can be reduced to the form \eqref{U_gate_general} by means of single-qubit $Z$ rotations applied from both sides of the operator, which can be implemented as virtual Z rotations in experiments~\cite{McKay2017}.

In addition to the SWAP-like interaction -- or $XX-YY$ interaction -- described by $\theta$ in Eq.~\eqref{U_gate_general}, the microwave drive creates a $ZZ$ interaction between computational states. This coupling leads to a finite $\zeta$ in  Eq.~(\ref{U_gate_general}), which cannot be changed to zero by local (single-qubit) operations. The $ZZ$ term has two distinct contributions: the static $ZZ$ coupling, which is caused by the repulsion between computational and noncomputational levels due to interaction (\ref{interaction-charge}), and the $ZZ$ coupling induced by the microwave drive used to perform the gate operation. While the effect of the static $ZZ$  coupling is relatively weak (the phase accumulation rate is slightly below 1 MHz for the parameters of Table~\ref{Table-params} with $J_C/h=200$ MHz)  and leads to a small contribution to $|\zeta| \ll \pi$ for  short gate durations, the microwave-induced contribution to $\zeta$ can be large, as demonstrated in Eq.~\eqref{eq:alpha}. Thus, one has to include the angle $\zeta$ in the definition of the target gate, Eq.~(\ref{U_gate_general}), to take into account this additional term caused by the drive.
As we demonstrate below for three choices of the mixing angle $\theta$, the value of $\zeta$ affects the equivalence class of the gate and its entangling properties.

\subsubsection{Mixing angle $\theta=\pi$}

First, we consider half a period of a resonant Rabi rotation, which corresponds to $\theta = \pi$ in Eq.~\eqref{U_gate_general}. If $\zeta=0$, the gate $U_{00-11}(\pi, 0)$ is bSWAP~\cite{Poletto2012}, which is a gate locally equivalent to iSWAP. On the other hand, $U_{00-11}(\pi, \pi)$ is locally equivalent to SWAP and thus does not generate  entanglement.

In general, two gates $U_{00-11}(\pi, \zeta)$ having different values of $\zeta$ in the interval between $0$ and $\pi$ are not locally equivalent. Each class of locally equivalent gates is characterized by special invariants $G_1=G_1'+iG_1''$ and $G_2$, and two gates are locally equivalent if and only if they have the same invariants~\cite{Makhlin2002,Zhang2003_pra}. We calculate them in Appendix~\ref{Sec:local-invariants} and find the following values for $U_{00-11}(\pi, \zeta)$:
\begin{equation}\label{eq:Gpi}
G_1'=-\sin^2 \frac{\zeta}{2},\quad 
G_1''=0,\quad 
G_2=\cos \zeta -2 \, .
\end{equation}

Another important property of a two-qubit gate is the entangling power ${\cal P}(\theta,\zeta)$~\cite{Zanardi2000,Ma2007}. It ranges
between ${\cal P}=0$ for 
$\hat{U}_{00-11}(\pi, \pi)$ (equivalent to a SWAP gate), and ${\cal P}=2/9$ for $\hat{U}_{00-11}(\pi, 0)$ (equivalent to an  iSWAP gate).  For arbitrary $\zeta$, the entangling power is
given by 
\begin{equation}\label{eq:Ppi}
{\cal P} (\theta=\pi, \zeta) = \frac{2}{9}\cos^2\frac{\zeta}{2},
\end{equation}
see Appendix ~\ref{Sec:entangling-power}.

\subsubsection{Mixing angle $\theta=\pi/2$}
When $\theta = \pi/2$, we also observe that the local equivalence class of $\hat{U}_{00-11}(\pi/2, \zeta)$ depends on $\zeta$. For this mixing angle, the local equivalence classes are characterized by the invariants
\begin{equation}\label{eq:Ghalfpi}
G_1'=\frac{\cos \zeta}{4},\quad 
G_1''=\frac{\sin \zeta}{4},\quad 
G_2=\cos \zeta \, .
\end{equation}
For example, $\hat{U}_{00-11}(\pi/2, 0)$ is $\sqrt{\rm bSWAP}$, which is locally equivalent to $\sqrt{\rm iSWAP}$~\cite{Poletto2012}, while $\hat{U}_{00-11}(\pi/2, \pi/2)$ and $\hat{U}_{00-11}(\pi/2, 3\pi/2)$ are equivalent to $\left(\sqrt{\rm SWAP}\right)^\dagger$ and $\sqrt{\rm SWAP}$, respectively, which are not locally equivalent to each other and to $\sqrt{\rm iSWAP}$. 

The entangling power of $\hat{U}_{00-11}(\pi/2, \zeta)$ is independent of $\zeta$ and is equal to ${\cal P}(\theta=\pi/2,\zeta)=1/6$.
Thus, a gate in the $\theta=\pi/2$  family is guaranteed  to be an entangling gate regardless of $\zeta$.  This is contrary to the case of $U_{00-11}(\theta =\pi, \zeta)$ gates, for which the entangling power varies according to Eq.~\eqref{eq:Ppi}.

The independence of the entangling power on values of $\zeta$ for $\theta = \pi/2$ makes this mixing angle an attractive choice in situations when the induced $ZZ$ coupling is hard to control. 
Another benefit of implementing a gate with $\theta=\pi/2$ vs a gate with $\theta = \pi$ is that the former gate  can be realized with any off-resonant two-photon Rabi oscillations as long as their contrast $V = \max_{t} P_{00 \to 11}(t)$ is at least $0.5$, while $\theta = \pi$ requires precise swapping of populations via half a period of a resonant Rabi rotation. For example, Fig.~\ref{Fig-rabi-osc}(d) demonstrates that it is possible to achieve $\theta = \pi/2$ by choosing half a period of the off-resonant Rabi oscillations with $V = 0.5$. 
 The drive detuning from the two-photon resonance, $\omega_d = (E_{11}-E_{00})/2\hbar$, changes the contrast $V$ and the period of  Rabi oscillations, which, in turn, affects the gate duration and $\zeta$ in $\hat{U}_{00-11}(\pi/2, \zeta)$, see, e.g., Eq.~\eqref{eq:alpha} for $\zeta$ at the resonant drive frequency, when $V=1$.  Thus, the detuning acts as an additional control, which  can be used either in the optimization procedure to improve gate performance when a specific $\zeta$ is not needed or in producing a gate with specific $\zeta$.

We also note that  $\hat{U}_{00-11}(\pi/2, \zeta)$ is sufficient to realize a  gate given by the unitary (\ref{U_gate_general}) with any mixing angle $\theta$ by combining two gates  $\hat{U}_{00-11}(\pi/2, \zeta)$ with single-qubit $Z$ rotations. Some of those $Z$ rotations can be substituted by a change of the microwave-drive phase $\gamma_d$ in the drive term~\eqref{drive} for one of the two-qubit gates. More details are given in Appendix~\ref{Sec:gates-decomposition}.  This decomposition is similar to the  decomposition in Ref.~\cite{Abrams2019} for $XY$ gates, which are excitation-preserving swapping gates activating coherent rotations in the $\{|01\rangle, |10\rangle\}$ subspace. 
We note that $\hat{U}_{00-11}(\pi/2, \zeta)$ is not a Clifford gate for any $\zeta$, which requires its characterization via the cross-entropy benchmarking~\cite{Boixo2018} rather than via randomized benchmarking~\cite{Magesan2012}. 

\subsubsection{Mixing angle $\theta=0$}
We finally consider the case when  $\theta=0$.  This gate occurs after a full two-photon Rabi oscillation with resonant, $\theta=\pi$, or off-resonant drive, e.g., $\theta=\pi/2$.  The gate is equivalent to the controlled-phase gate with 
the invariants determined by $\zeta$ and coinciding with invariants of the controlled-phase gate:
\begin{equation}\label{eq:Gzero}
G_1'=\cos^2 \frac{\zeta}{2},\quad 
G_1''=0,\quad 
G_2=\cos \zeta + 2 \, .
\end{equation}
The entangling power is also $\zeta$ dependent:
\begin{equation}
{\cal P} (\theta=0, \zeta) = \frac{2}{9}\sin^2\frac{\zeta}{2}.
\end{equation}

\subsection{Simulated coherent gate fidelity}\label{Sec:gates-simulations}

Here we demonstrate via numerical simulations that a fast and high-fidelity gate that mixes states $\ket{00}$ and $\ket{11}$ by means of a monochromatic microwave drive is possible. We focus on the gate operation for the mixing angle $\theta = \pi/2$ and calculate the gate fidelity to an ideal unitary $\hat{U}(\pi/2, \zeta)$ with a suitable choice of $\zeta$ angle. We start from a detailed analysis of coherent gate dynamics and discuss incoherent error in Sec.~\ref{Sec:incoherent-error}.  As in Sec.~\ref{Sec:Rabi-numerics}, we perform simulations for qubit parameters shown in Table~\ref{Table-params} and for fixed and equal drive couplings in Eq.~\eqref{drive}, $\eta_A = \eta_B=1$. 

For a given gate duration $t_{\rm gate}$, we use the Gaussian pulse shape with the rising edge given by Eq.~\eqref{pulse-shape}, where $t_{\rm rise} = t_{\rm gate}/2$ and $\sigma = t_{\rm rise}/2$. After solving the Schr\"odinger equation given by the time-dependent Hamiltonian \eqref{Hamiltonian-two-qubit} for four initial states in $\{\ket{00}, \ket{01}, \ket{10}, \ket{11}\}$, we obtain the propagator describing the evolution of computational states in a larger Hilbert space also containing higher-energy levels.
 Projecting this operator into the computational subspace yields the simulated gate operator $\hat{U}_{\rm sim}$. By adding single-qubit $Z$ rotations before and after the gate, we adjust phases of relevant matrix elements of $\hat{U}_{\rm sim}$ to compare it with $\hat{U}_{00-11}(\pi/2, \zeta)$; more details are given in Appendix~\ref{Sec:error-budget}.
 For the target unitary $\hat{U}_{00-11}(\pi/2, \zeta)$, we choose
$\zeta = \varphi_{01, 01} + \varphi_{10, 10} - \varphi_{00, 00} - \varphi_{11, 11}$, where $\varphi_{kl, kl} = {\rm arg}\bra{kl} \hat{U}_{\rm sim}\ket{kl}$ is the diagonal-matrix-element phase of $\hat{U}_{\rm sim}$, so a new value of $\zeta$ is chosen each time a new $\hat{U}_{\rm sim}$ is computed.  Denoting the simulated gate operator after the application of $Z$ rotations as $\hat{U}$, we use the standard expression for the two-qubit-gate fidelity~\cite{Pedersen2007}:
\begin{equation}\label{fidelity_unitary}
    F = \frac{{\rm Tr} \left(\hat{U}^\dagger \hat{U}\right) + \left|{\rm Tr}\left[\hat{U}^\dagger \hat{U}_{00-11}(\pi/2, \zeta)\right]\right|^2}{20}\,.
\end{equation}
Using this metric, we optimize the coherent gate error over the drive frequency and amplitude. We analyze the dependence of $F$ on the total gate duration $t_{\rm gate}$ and the interaction strength $J_C$.

\begin{figure}[t]
\includegraphics[width=\columnwidth]{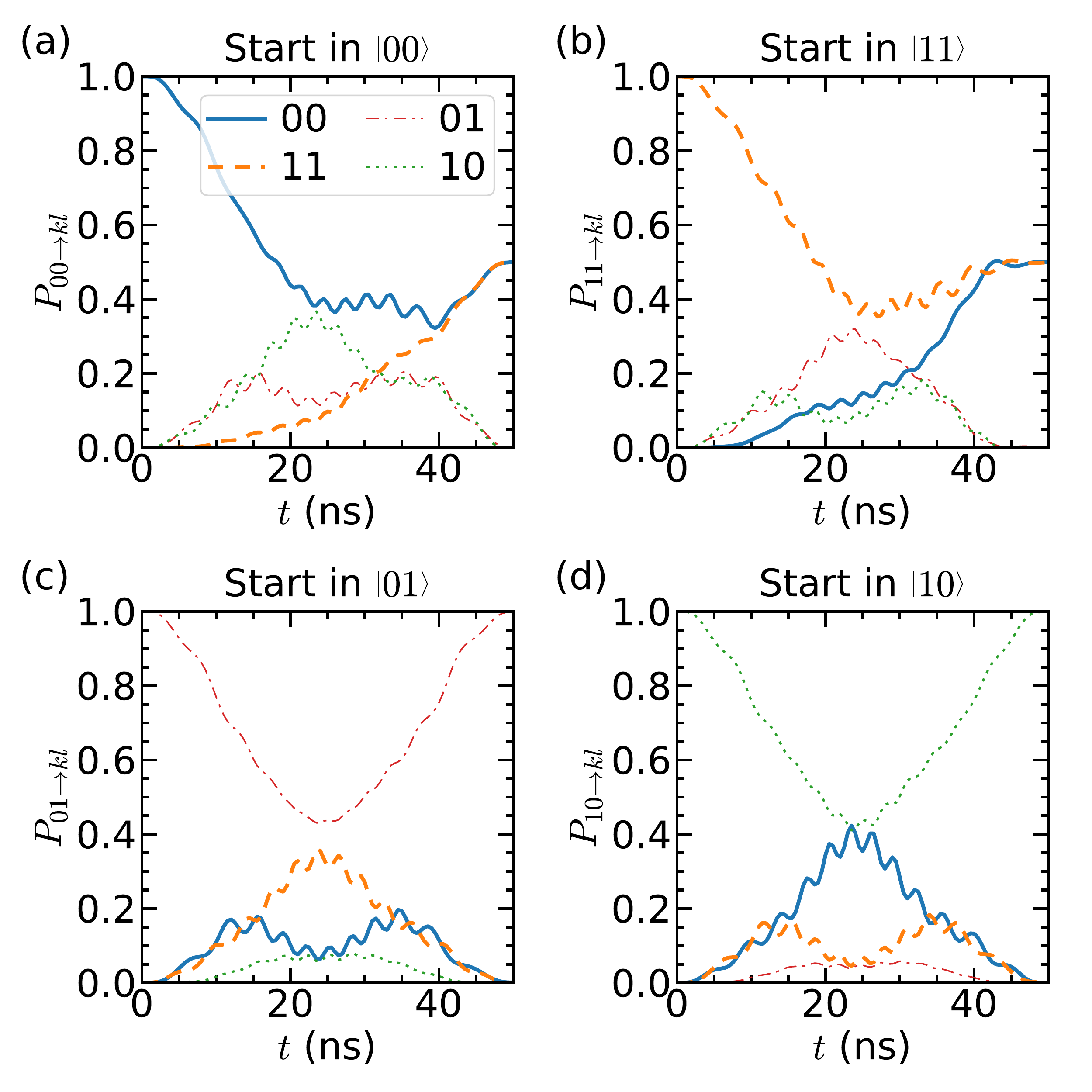}\caption{
Unitary time evolution of the populations during the gate at $J_C/h = 200$ MHz for four initial computational states. The gate is optimized over the drive amplitude and frequency at fixed $t_{\rm gate} = 50$ ns for the target operator  $\hat{U}_{00-11}(\pi/2, \zeta)$. The gate fidelity is $F \approx 0.99905$ and $\zeta \approx 1.02 \pi$. 
}\label{Fig-tdomain}
\end{figure}

\begin{figure}[t]
    \centering
    \includegraphics[width=\columnwidth]{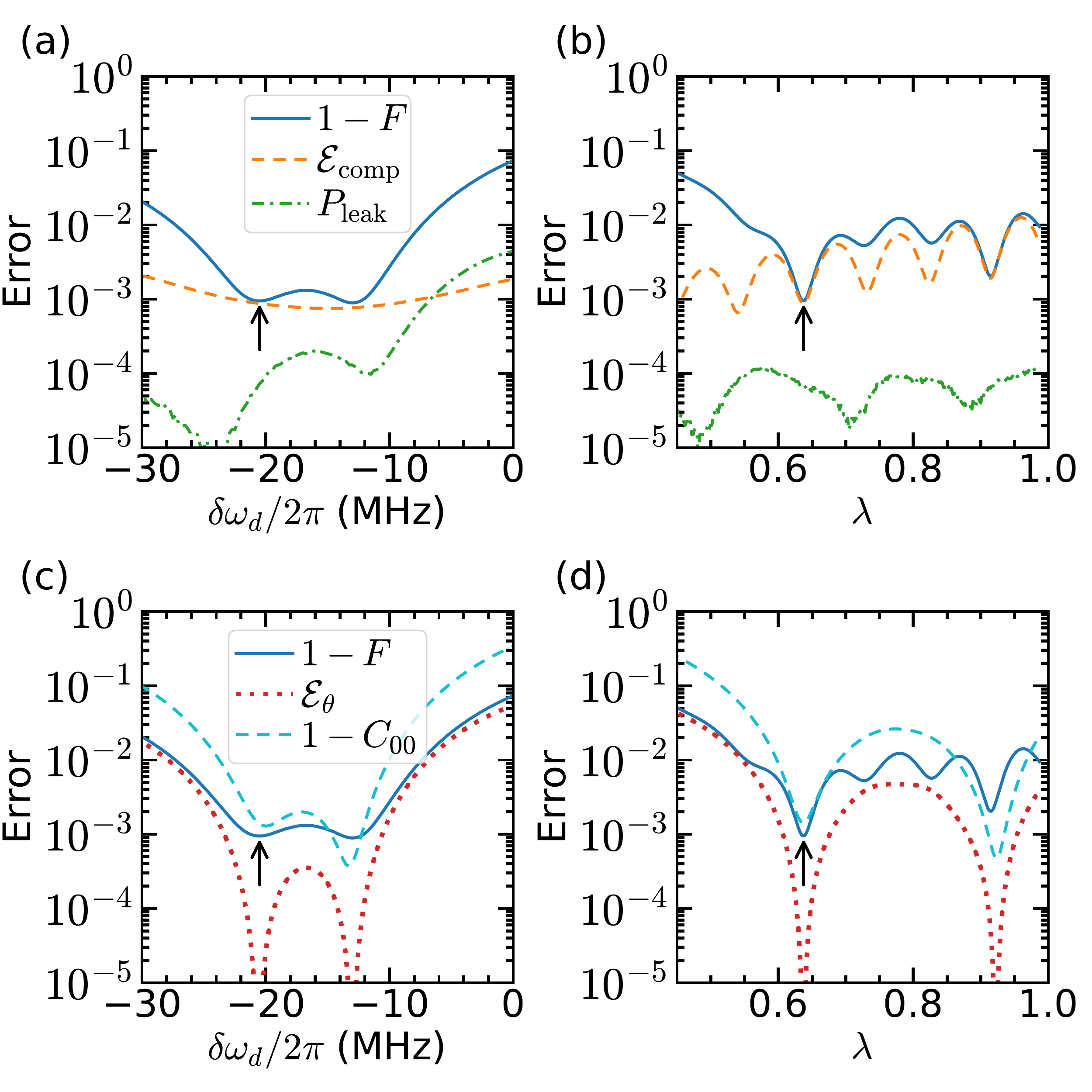}
    \caption{ Total coherent gate error  $1-F$ (solid lines) and contributions to $1-F$ vs $\delta\omega_d = \omega_d -\bar\omega$  at fixed $\lambda$ (left panels) and vs $\lambda$ at fixed  $\omega_d$ (right panels) for the same parameters as in Fig.~\ref{Fig-tdomain}. The drive amplitude is the amplitude averaged over the pulse duration. The optimal values of $\delta\omega_d$ and $\lambda$, used in simulations of Fig.~\ref{Fig-tdomain}, are marked by vertical arrows. (a), (b) Contributions to $1-F$ linear in population errors: the error due to incorrect populations of computational states $\mathcal{E}_{\rm comp}$ (dashed lines) and the average leakage probability to noncomputational states $P_{\rm leak}$ (dash-dot lines); see Eqs.~\eqref{error-comp} and \eqref{error-leakage}. 
  (c), (d) Mixing-angle error between $\ket{00}$ and $\ket{11}$ (dotted lines), which is quadratic in the angle error,  and  concurrence error after the gate operation on $|{00}\rangle$  (dashed lines); see Eqs.~\eqref{error-mixing}  and \eqref{concurrence}.}
    \label{Fig-omegadep}
\end{figure}

As an example, we first illustrate the gate operation in time domain in Fig.~\ref{Fig-tdomain} for such an optimized gate with $t_{\rm gate} = 50$ ns and $J_C/h= 200$ MHz. We show transition probabilities $P_{kl \to k'l'}$ vs time for all 16 pairs of initial and final states formed from $\{\ket{00}, \ket{01}, \ket{10}, \ket{11}\}$. The simulated gate is found to be in the equivalence class determined by $\theta=\pi/2$ and $\zeta \approx 1.02\pi$ with the gate fidelity being $F\approx 0.99905$. The gate duration of $t_{\rm gate} = 50$ ns is  the shortest time for which $1-F < 0.001$ for the interaction strength used in this simulation.

While the time evolution of state populations for the optimized gate of Fig.~\ref{Fig-tdomain} exhibits multiple fluctuations and large transient excitations of one-photon processes, the gate performance is actually very robust to calibration errors.
We illustrate this statement in Figs.~\ref{Fig-omegadep}(a) and \ref{Fig-omegadep}(b), where we study gate properties around the optimal point by changing the drive frequency and amplitude. We note that $1-F$ is below $0.001$ in the frequency interval greater than 1 MHz, see the solid line near the vertical arrow in Fig.~\ref{Fig-omegadep}(a). Longer gate durations can be chosen for which this interval is even wider.  

We analyze various coherent contributions to the gate error in Fig.~\ref{Fig-omegadep} with more details given in Appendix~\ref{Sec:error-budget}. For the ideal gate operation $\hat{U}_{00-11}(\pi/2, \zeta)$, the transition probabilities $P_{kl \to k'l'}$ are either zero, $1/2$, or 1, while the actual gate operation contains errors in those probabilities. When errors are small, $1-F$ is well approximated by the sum of two distinct contributions that are linear in  $P_{kl \to k'l'}$. The first contribution, $\mathcal{E}_{\rm comp}$, is determined by those $P_{kl \to k'l'}$ for which $|k'l'\rangle$ is in the computational subspace; see Eq.~\eqref{error-comp}. The second contribution, $P_{\rm leak}$, is the leakage error given by the average probability to end up outside of the computational subspace, see Eq.~\eqref{error-leakage}. We show these two contributions by dashed and dash-dot lines in Fig.~\ref{Fig-omegadep}(a) and \ref{Fig-omegadep}(b). We find that at the optimal point, the coherent gate error $1-F$ is determined by the error in the computational subspace $\mathcal{E}_{\rm comp}$. Leakage errors are below $0.0001$ at the optimal point. A more detailed analysis (not shown here) indicates that the remaining leakage errors are mostly coming from excitations of the second excited states of fluxonium circuits via, e.g., transitions $\ket{10}-\ket{20}$ and $\ket{11}- \ket {12}$. 

While $\mathcal{E}_{\rm comp}$ and $P_{\rm leak}$ are sufficient to explain the behavior of $1-F$ near its minimum, other error mechanisms become  dominant away from the optimal point. For instance, we find a large contribution from the mixing error $\mathcal{E}_\theta$. This error is determined by the differences $|P_{00\to 00} - P_{00\to 11}|$ and $|P_{11\to 00} - P_{11\to 11}|$ and is quadratic in them. Essentially, it can be thought of as the error in the mixing angle $\theta$ in $\hat{U}_{00-11}(\theta, \zeta)$. We show this error by the dotted line in Figs.~\ref{Fig-omegadep}(c) and \ref{Fig-omegadep}(d), which explains well the behavior of $1-F$ far away from the optimal point.

In addition to the gate fidelity, we calculate the concurrence $C_{00}$ (see Ref.~\cite{Wooters1998}) of a state vector starting in $\ket{00}$ after the application of the gate, see Appendix~\ref{Sec:error-budget}. It is shown by the dashed lines in Figs.~\ref{Fig-omegadep}(c) and \ref{Fig-omegadep}(d).  When $C_{00}=1$, the state is maximally entangled and the gate is thus a perfect entangler.
 While the mixing error has two almost symmetric minima, which are sharp and deep, the two minima of the concurrence error are asymmetric. This is explained by the dependence of the concurrence not only on the mixing of  states $|00\rangle$ and $|11\rangle$, but also on both amplitudes and phases of states $\ket{01}$ and $\ket{10}$, see Eq.~\eqref{concurrence}. The corresponding contributions coming from $\ket{01}$ and $\ket{10}$ have opposite signs in the left and right minima. 

\begin{figure}[t]
\includegraphics[width=\columnwidth]{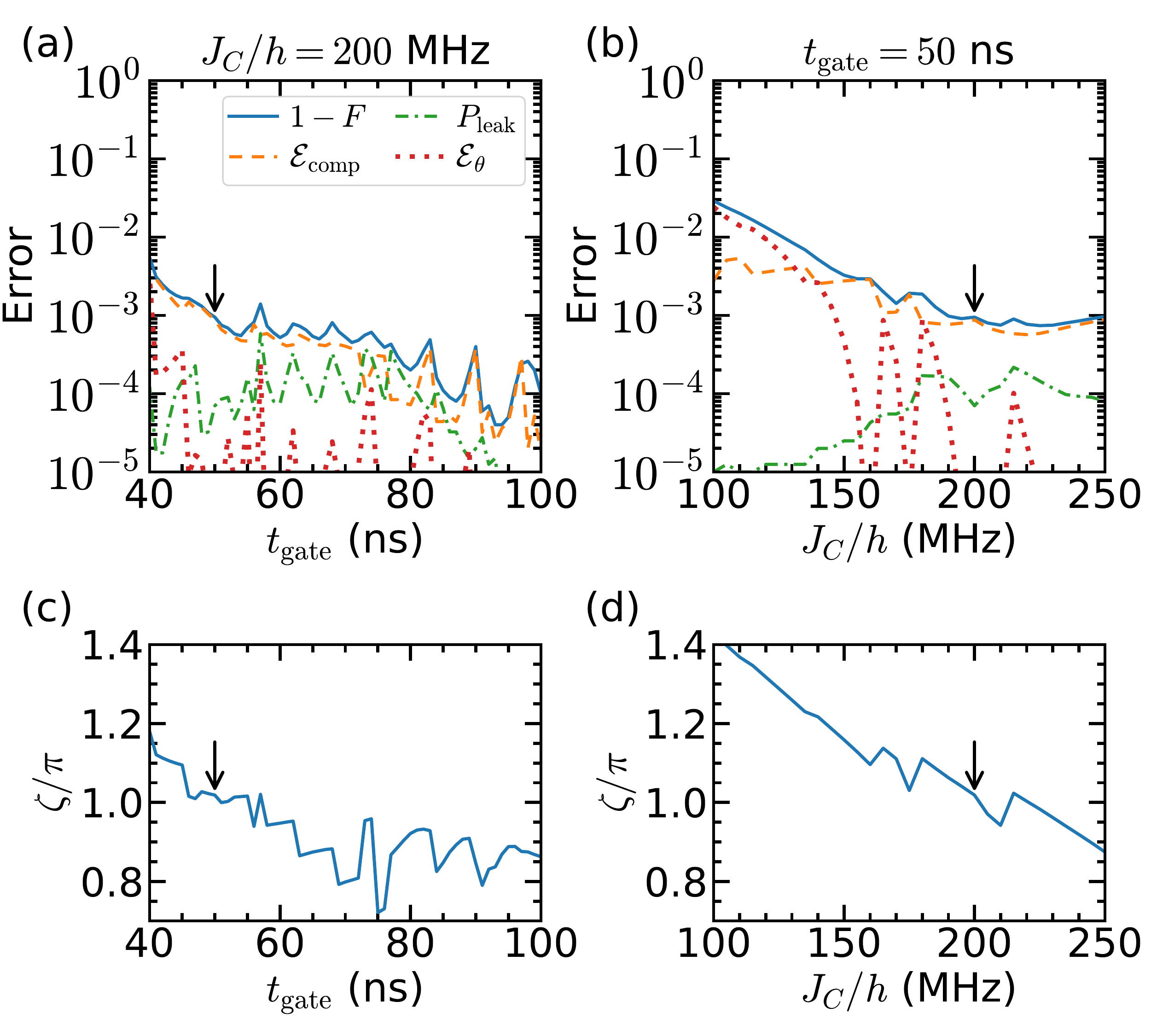}\caption{
(a), (b) Optimized coherent gate error and error budget vs total gate duration $t_{\rm gate}$ at $J_C/h = 200$ MHz (a) and vs $J_C/h$ at $t_{\rm gate} = 50$ ns (b). Line styles follow the convention of Fig.~\ref{Fig-omegadep}. (c), (d) Parameter $\zeta$ determining the local-equivalence class  [see Eq.~\eqref{U_gate_general}] of the same optimized gates as in the corresponding top panels. Vertical arrows in all four panels indicate the parameters of Fig.~\ref{Fig-tdomain}.
}\label{Fig-fidelity}
\end{figure}

In Fig.~\ref{Fig-fidelity}, we show coherent gate error and its budget as a function of gate duration $t_{\rm gate}$ and the interaction strength $J_C$. For $J_C/h=200$ MHz, we observe that the coherent error can easily go below $10^{-4}$ for a gate duration shorter than 100 ns. In the bottom panels, we show parameter $\zeta$ that determines the equivalence class of the gate. Its tendency to decrease with increasing $t_{\rm gate}$ or $J_C$ is explained by the contribution to $\zeta$ coming from the static $ZZ$ coupling. The effect of static $ZZ$ grows with the gate duration and interaction strength.  Finally, we note that combining two $\hat{U}_{00-11}(\pi/2, \zeta)$ gates, we can obtain a controlled-phase gate with phase $2\zeta$.

\subsection{Incoherent error}\label{Sec:incoherent-error}

In this section, we discuss how qubit decoherence  affects the gate error.  We consider relaxation and dephasing of only $\ket{0_\alpha}-\ket{1_\alpha}$ and $\ket{1_\alpha}-\ket{2_\alpha}$ transitions since the qubit excitation probability above its second excited state is very small. For example, for the parameters of Fig.~\ref{Fig-tdomain}, the maximum and average populations of the third excited states are at most 0.5\% and 0.1\%, respectively, so qubits spend at most 0.05 ns in their third excited states during 50 ns of gate operation.  For states $\ket{2_\alpha}$, these numbers are about an order of magnitude larger (5\%, 1.2\%, and 0.6 ns) and are thus not large either, but  may still result in an important contribution to the gate error since the coherence time of $\ket{1_\alpha}-\ket{2_\alpha}$ is often significantly shorter than the coherence time of the computational subspace. A small population of state $\ket{2_\alpha}$ is consistent with a small $\lambda_{1-2}$, defined in Eq.~\eqref{lambda_1_2}.

We simulate the gate operation in the presence of decoherence for optimal pulse parameters found in simulation of unitary dynamics in Sec.~\ref{Sec:gates-simulations}. The evolution of a density matrix $\rho$ is described by the Lindblad master equation
\begin{equation}\label{Lindblad}
    \frac{d\hat{\rho}}{dt} = -\frac{i}{\hbar}\left[\hat{H}, \hat{\rho}\right] + \sum_k \left[\hat{L}_k \hat{\rho}\hat{L}_k^\dagger - \frac 12 \left(\hat{L}_k^\dagger \hat{L}_k \hat{\rho} + \hat{\rho} \hat{L}_k^\dagger \hat{L}_k\right)\right]\,,
\end{equation}
where we use eight collapse operators $\hat{L}_k$ corresponding to relaxation and pure dephasing in two transitions in both qubits. For qubit A transitions, we form them as follows:
\begin{subequations}
 \begin{gather}
     \hat{L}_1^{0-1, A} = \sqrt{\frac{1}{T^{0-1, A}_1}}\sum_k \ket{0k} \bra{1k}\,, \\
     \hat{L}_\varphi^{0-1, A} = \sqrt{\frac{2}{T^{0-1, A}_\varphi}}\sum_k \ket{0k} \bra{0k}\,,\\
     \hat{L}_1^{1-2, A} = \sqrt{\frac{1}{T^{1-2, A}_1}}\sum_k \ket{1k} \bra{2k}\,, \\
     \hat{L}_\varphi^{1-2, A} = \sqrt{\frac{2}{T^{1-2, A}_\varphi}}\sum_k \ket{2k} \bra{2k}\,.
 \end{gather}
\end{subequations}
Here $T^{k-l, A}_1$ and $T^{k-l, A}_\varphi$ are the relaxation and pure dephasing times of the $\ket{k_A}-\ket{l_A}$ transition of qubit A.
The collapse operators for qubit B transitions are formed in a similar way with time parameters $T^{k-l, B}_1$ and $T^{k-l, B}_\varphi$.
This approach to describe relevant noise sources is sufficient for our purpose of providing a crude estimate. 

Using these collapse operators, we perform numerical quantum process tomography. We simulate the superoperator  describing the evolution of density matrices  corresponding to master equation~\eqref{Lindblad}, project the operator into the computational subspace, and use it to find the $16 \times 16$ $\chi$ matrix $\chi_{\rm real}$ describing the quantum process. We find the ideal $\chi$ matrix $\chi_{\rm ideal}$ using $\hat{U}_{00-11}(\pi/2, \zeta)$ in Eq.~\eqref{fidelity_unitary} modified by single-qubit $Z$ rotations used to obtain $\hat{U}$ from $\hat{U}_{\rm sim}$ in unitary simulations; see the text above Eq.~\eqref{fidelity_unitary}. We then use
\begin{equation}\label{fidelity_nonunitary}
    F = \frac{4{\rm Tr}(\chi^\dagger_{\rm real}\chi_{\rm ideal}) + {\rm Tr}(\chi_{\rm real})}{5}\,,
\end{equation}
which establishes a relation between the gate and process fidelities, where the latter is given by ${\rm Tr}(\chi^\dagger_{\rm real}\chi_{\rm ideal})$~\cite{Chow2009}.

\begin{figure}[t]
\includegraphics[width=\columnwidth]{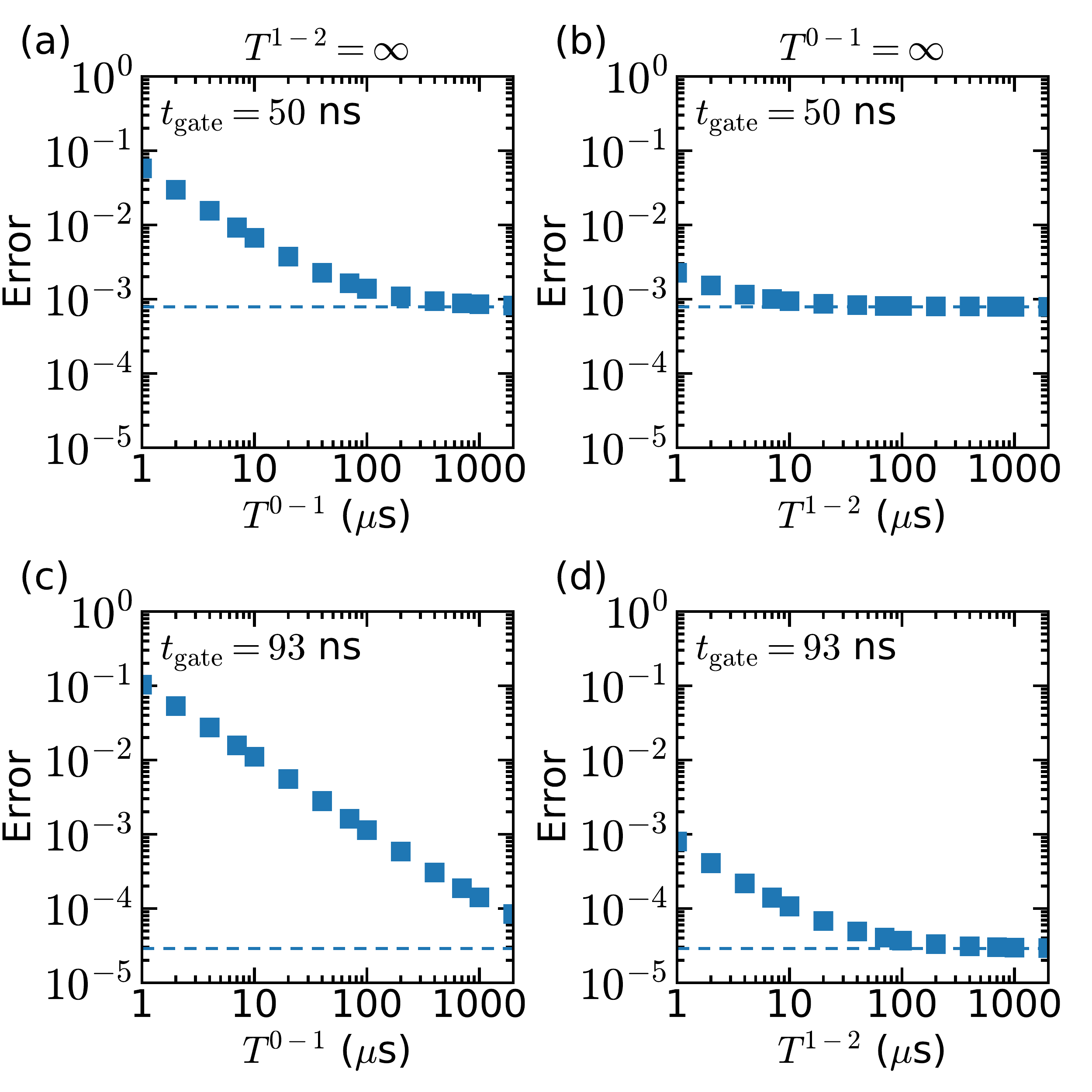}\caption{Gate error (squares) vs relaxation and dephasing times for the parameters of  Fig.~\ref{Fig-fidelity}(a) corresponding to $t_{\rm gate}=50$ ns (a), (b) and $t_{\rm gate} = 93$ ns (c), (d).  Relaxation and dephasing are accounted for either only in $\ket{0_\alpha}-\ket{1_\alpha}$ single-qubit transitions (a), (c) or only in $\ket{1_\alpha}-\ket{2_\alpha}$ transitions (b), (d) with $T_1$ and $T_2$ times being the same for both qubits and equal to each other. Dashed lines show the coherent error.}\label{Fig-dissip}
\end{figure}

Using this approach, we study how gate error depends on relaxation and dephasing times. For each transition $\ket{k_\alpha}-\ket{l_\alpha}$,  we assume that its relaxation ($T^{k-l, \alpha}_1$) and coherence ($T^{k-l, \alpha}_2$) times are the same, so its pure dephasing time is $T^{k-l, \alpha}_\varphi = 2T^{k-l, \alpha}_1$. We also assume that these times are the same for both qubits, but different for the two transitions, so we use two lifetime parameters: $T^{0-1}$ for relaxation and dephasing of the computational transitions of both qubits and $T^{1-2}$ for the $\ket{1_\alpha}-\ket{2_\alpha}$ transitions.
In the top panels of Fig.~\ref{Fig-dissip}, we show the gate error calculated for Fig.~\ref{Fig-fidelity}(a) parameters at $t_{\rm gate} = 50$ ns, which are marked by vertical arrows in Fig.~\ref{Fig-fidelity} and were also discussed in Figs.~\ref{Fig-tdomain} and \ref{Fig-omegadep}. In the bottom panels of Fig.~\ref{Fig-dissip}, we consider Fig.~\ref{Fig-fidelity}(a) parameters at $t_{\rm gate} = 93$ ns, which is a local minimum of the coherent error. We study separately the effects of decoherence of $\ket{0_\alpha}-\ket{1_\alpha}$ and $\ket{1_\alpha}-\ket{2_\alpha}$ transitions. Thus, the left panels of Fig.~\ref{Fig-dissip} show $1-F$ vs $T^{0-1}$ assuming that $T^{1-2}=\infty$ and the right panels discuss $T^{0-1}=\infty$ and finite $T^{1-2}$. Horizontal dashed lines show the coherent error, which was calculated in the previous section, so the difference between symbols and lines is the incoherent contribution coming from either $\ket{0_\alpha}-\ket{1_\alpha}$ or $\ket{1_\alpha}-\ket{2_\alpha}$ transitions. The total incoherent  error is approximately given by the sum of incoherent errors in the left and right panels. 

Figure~\ref{Fig-dissip} demonstrates that the contribution to the incoherent error coming from $\ket{1_\alpha}-\ket{2_\alpha}$ transitions is much less important than that coming from the relaxation and dephasing in the computational subspace. This is consistent with small average occupations of the second excited states of fluxoniums during the gate operation. We observe that even a very short $T^{1-2} \sim 1$ $\mu$s results in only about $10^{-3}$ contribution to the gate error, while $T^{1-2}\sim 20$ $\mu$s is sufficient to contribute less than $0.5 \times 10^{-4}$ at $t_{\rm gate} = 93$ ns. Several dozens of microseconds for a coherence time of a transition with frequency in the gigahertz scale is common in modern state-of-the-art superconducting qubits~\cite{Kjaergaard2020_review} with the best lifetimes exceeding 100 $\mu$s~\cite{Nersisyan2019, Place2021}. Therefore, we do not expect  the coherence time of the $\ket{1_\alpha}-\ket{2_\alpha}$ transition to be a limiting factor for the proposed gate. The contribution from decoherence of the computational transitions is more important. We find that  $T^{0-1}\sim 100$ $\mu$s is generally sufficient to bring the gate error below $10^{-3}$, while the $10^{-4}$ threshold requires $T^{0-1} > 1$ ms. We note that the best fluxonium devices have recently demonstrated lifetimes of 1 ms, although at lower transition frequencies than those discussed here~\cite{Somoroff2021}. Nevertheless, we do not see any fundamental obstacles in achieving a millisecond lifetime of the fluxonium with 1 GHz transition, which paves the way towards $10^{-4}$ gate errors.

\section{Discussion and Conclusions}\label{Sec:conclusions}

We demonstrated that fast high-fidelity microwave-activated two-qubit gates are possible in fluxonium circuits when the system state remains entirely in the low-energy computational subspace.   Despite a relatively weak effect of capacitive interaction between fluxoniums on the computational subspace, the gate time can still be as short as 50 ns due to the strong anharmonicity of the fluxonium spectrum.  The anharmonicity typically limits the intensity of the microwave drive.  We demonstrated that the microwave amplitude could be large for the proposed two-photon gate without causing noticeable leakage of the state outside of the computational subspace during the pulse, minimizing the effect of decoherence on the gate fidelity. The required  amplitude is about 10-20 times larger than its value in schemes utilizing noncomputational levels~\cite{Ficheux2021, Xiong2021} and in single-qubit operations. This strength is on par with the cross-resonance gate, which is activated by microwave fields with resonance Rabi frequencies up to hundreds of megahertz~\cite{Chow2011} and which has techniques to mitigate cross-talk and spectator errors in transmon processors~\cite{Sheldon2016b, Sundaresan2020}. Strong anharmonicity of the fluxonium and an extra freedom in choosing the qubit frequency will likely make mitigation of these errors even more successful in a fluxonium-based processor.

At a weak drive power, the rate of two-photon transitions is quadratic in the drive amplitude. In this case, the gate would be prohibitively long if the drive were chosen so that single-photon transitions between subspaces $\{\ket{00},\ket{11}\}$ and $\{\ket{01},\ket{10}\}$ were strongly suppressed. We demonstrated that unintended single-photon transitions between those subspaces could be reduced even for a strong drive by fine tuning the pulse amplitude and frequency together with the gate duration. As a result, the microwave pulse only mixes states $\ket{00}$ and $\ket{11}$.  We focused on the half-mixing angle $\theta=\pi/2$ for which the entangling power is independent of the phase shift due to the ZZ interaction, which guarantees that the gate is entangling without  any control of the Stark-induced phase accumulation.  We also note that the half-mixing gate is shorter and is often more robust to pulse imperfections. 

In conclusion, we considered a two-qubit gate that is well suited for existing fluxonium devices and is ready to be implemented. 
The proposed scheme is very generic and can also be realized as a two-color scheme with two microwave drives at two different frequencies $\omega_{d1}$ and $\omega_{d2}$ satisfying $\omega_{d1} + \omega_{d2} = \omega_{00-11}$, which provides additional controls to reduce errors. The gate works for fluxoniums parked at the sweet spots of their maximal coherence and does not require any additional hardware beyond microwave control lines necessary to activate single-qubit gates.

\begin{acknowledgements}

We would like to thank Mark Dykman, Ivan Pechenezhskiy, Haonan Xiong, and Long Nguyen for fruitful discussions. We acknowledge the support from NSF PFC at JQI and ARO-LPS HiPS program (grant No. W911NF-18-1-0146). V.E.M. and M.G.V acknowledge the Faculty Research Award from Google and fruitful conversations with the members of the Google Quantum AI team.  We used the QuTiP software package~\cite{Johansson2012, Johansson2013}  and performed computations using resources and assistance of the UW-Madison Center For High Throughput Computing (CHTC) in the Department of Computer Sciences. The CHTC is supported by UW-Madison, the Advanced Computing Initiative, the Wisconsin Alumni Research Foundation, the Wisconsin Institutes for Discovery, and the National Science Foundation.

\end{acknowledgements}

\appendix

\section{Local invariants}\label{Sec:local-invariants}
Here we calculate the invariants~\cite{Makhlin2002, Zhang2003_pra} for the gates described by Eq.~(\ref{U_gate_general}) and demonstrate that the Stark shift $\zeta$ makes the gates with different $\zeta$ nonequivalent to each other. 

To calculate the local invariants, we express the gate operator in the Bell basis defined by the transformation operator
$$
Q = \frac{1}{\sqrt{2}}
\begin{pmatrix}
 1 & 0 & 0 & i \\
 0 & i & 1 & 0 \\
 0 & i & -1 & 0 \\
 1 & 0 & 0 & -i
\end{pmatrix}\,.
$$
We define 
$U_B = Q^\dagger U_{00-11}(\theta, \zeta) Q $
and 
$
m(U_{00-11}) = U_B^T U_B 
$.
The local invariants~\cite{Zhang2003_pra} are given by:
\begin{subequations}
 \begin{equation}\label{G_1_invariant}
  G_1 = \frac{{\rm tr}^2\left[m(U_{{00-11}})\right]}{16\det U_{{00-11}}} = \frac{e^{-i\zeta}}{4}(e^{i\zeta}+\cos\theta)^2
 \end{equation}
and
\begin{equation}\label{G_2_invariant}
 G_2 = \frac{{\rm tr}^2\left[m(U_{{00-11}})\right] - {\rm tr}\left[m^2(U_{{00-11}})\right]}{4\det U_{{00-11}}} = 2\cos\theta+\cos\zeta \,,
\end{equation}
\end{subequations}
where $G_1 = G_1'+iG_1''$ is a complex number and $G_2$ is real.
We obtain $G_1$ and $G_2$ for arbitrary $\zeta$ and  three choices of $\theta = 0$, $ \pi/2$ and $\pi$, given in Eqs.~\eqref{eq:Gpi},  \eqref{eq:Ghalfpi}, and \eqref{eq:Gzero}.

\section{Entangling power}\label{Sec:entangling-power}

Here we provide expressions for calculations of the entangling power ${\cal P}$ of a two-qubit  operator (\ref{U_gate_general})~\cite{Zanardi2000}. We use the algebraic technique of Ref.~\cite{Ma2007} that defines 
the entangling power as
\begin{equation}\label{entangling_power}
{\cal P}(U) = \frac{4}{9}\left[E(U) + E(US_{12}) - E(S_{12})\right]\,,
\end{equation}
where $E(U)$ is the operator entanglement (linear entropy) of $U$:
$$
E(U) = 1 - \frac{1}{16}{\rm Tr}\left[U^R (U^R)^\dagger U^R (U^R)^\dagger\right]\,,
$$
the matrix $U^R$ is obtained from $U$ by realignment:
$$
(U^R)_{ij, kl} = U_{ik, jl}\,,
$$
and $S_{12}$ is the swapping operator
$$
S_{12} = \sum_{ij} |ij\rangle \langle ji | = 
\begin{pmatrix}
 1 & 0 & 0 & 0 \\
 0 & 0 & 1 & 0 \\
 0 & 1 & 0 & 0 \\
 0 & 0 & 0 & 1
\end{pmatrix}\,.
$$

We take $U(\theta,\zeta)$ in the form of Eq.~\eqref{U_gate_general} and find that
\begin{equation}
{\cal P}(\theta,\zeta) = \frac{1}{36}\left(
5-4\cos\zeta\cos\theta-\cos2\theta
\right).
\end{equation}
We note that  the entangling power is independent of the Stark shift $\zeta$ only for $\theta=\pi/2$.

\section{Gates decomposition} \label{Sec:gates-decomposition}

Here we show how an operation $\hat{U}_{00-11}(\theta, \zeta)$, which is given by Eq.~\eqref{U_gate_general}, can be decomposed into two operations $\hat{U}_{00-11}(\pi/2, \zeta/2)$ and single-qubit $Z$ rotations. 
This decomposition has the form
\begin{multline}\label{U_gate_decomposition_any_swapping}
    \hat{U}_{00-11}(\theta, \zeta) =  \hat{U}_{Z}\left(-\frac{\theta}{4}\right)\hat{{U}}'_{00-11}\left(\frac{\pi}{2}, \frac{\zeta}{2}, \frac{\theta}2-\frac{\pi}{2}\right)
    \\
    \times \hat{{U}}'_{00-11}\left(\frac{\pi}{2}, \frac{\zeta}{2}, -\frac{\theta}2+\frac{\pi}{2}\right) \hat{U}_{Z}\left(-\frac{\theta}{4}\right)\,,
\end{multline}
where
\begin{equation}
    \hat{U}_Z(\nu) = \hat{Z}_\nu^A \otimes \hat{Z}_\nu^B = {\rm diag}\left(e^{-i\nu}, 1, 1, e^{i\nu}\right)
\end{equation}
is the tensor product of two single-qubit $Z$ rotations $\hat{Z}_{\nu} = {\rm diag}\left(e^{-i\nu/2}, e^{i\nu/2}\right)$ of qubits $A$ and $B$ and
\begin{equation}\label{U_gate_adding_off_diagonal}
    \hat{U}'_{00-11}\left(\frac{\pi}{2}, \frac{\zeta}{2}, \gamma\right) = \hat{U}_Z\left(\frac{\gamma}{2}\right)\hat{U}_{00-11}\left(\frac{\pi}{2}, \frac{\zeta}{2}\right) \hat{U}_Z\left(-\frac{\gamma}{2}\right)\,.
\end{equation}
The last gate operation differs from $\hat{U}_{00-11}({\pi}/{2}, \zeta/2)$  by having additional phase factors $e^{i\gamma}$ and $e^{-i\gamma}$ in the off-diagonal matrix elements of the $\{|00\rangle, |11\rangle\}$ subspace. In addition to performing $Z$ rotations as in Eq.~\eqref{U_gate_adding_off_diagonal}, this can be achieved by adjusting the microwave-drive phase $\gamma_d$ in Eq.~\eqref{drive}: a change $\delta\gamma_d$ in that phase results in the change of $2\delta\gamma_d$ in $\gamma$~\cite{Poletto2012}. Essentially, this is equivalent to virtual $Z$ rotations, which amount to a change of the reference frame by changing phases of microwave pulses~\cite{McKay2017}. The decomposition \eqref{U_gate_decomposition_any_swapping} is reminiscent of a similar decomposition in Ref.~\cite{Abrams2019} for $XY$ gates, which are swapping gates in the $\{|01\rangle, |10\rangle\}$ subspace.

\section{Coherent gate fidelity and error budget} \label{Sec:error-budget}

In the ideal gate operation $\hat{U}_{00-11}(\pi/2, \zeta)$ with $\pi/2$ mixing angle, transition probabilities $P_{kl\to k'l'}$ are either 0, 1, or 1/2. Here we assume small deviations $\epsilon_{k'l',kl}$  from such values and calculate their contribution to the gate error. 
By using single-qubit $Z$ rotations from both sides, we can write a realistic gate operator projected into the computational subspace as
\begin{widetext}
\begin{equation}\label{U_real}
    \hat{U}_{\rm real} = 
    \begin{pmatrix}
    \sqrt{\frac 12 - \epsilon_{00, 00}} & \sqrt{\epsilon_{00, 01}}\;e^{i\varphi_{00, 01}} & \sqrt{\epsilon_{00, 10}}\;e^{i\varphi_{00, 10}} & \sqrt{\frac 12 - \epsilon_{00, 11}}\left(-ie^{i\beta/2 }\right) \\
    \sqrt{\epsilon_{01, 00}}\;e^{i\varphi_{01, 00}} & \sqrt{1 - \epsilon_{01, 01}}\; e^{i\zeta/2} & \sqrt{\epsilon_{01, 10}}\;e^{i\varphi_{01, 10}} &\sqrt{\epsilon_{01, 11}}\;e^{i\varphi_{01, 11}} \\
    \sqrt{\epsilon_{10, 00}}\;e^{i\varphi_{10, 00}} & \sqrt{\epsilon_{10, 01}}\;e^{i\varphi_{10, 01}} & \sqrt{1 - \epsilon_{10, 10}}\; e^{i\zeta/2} & \sqrt{\epsilon_{10, 11}}\;e^{i\varphi_{10, 11}} \\
   \sqrt{\frac 12 - \epsilon_{11, 00}}\left(-ie^{i\beta/2 }\right) & \sqrt{\epsilon_{11, 01}}\;e^{i\varphi_{11, 01}} & \sqrt{\epsilon_{11, 10}}\;e^{i\varphi_{11, 10}} & \sqrt{\frac 12 - \epsilon_{11, 11}}
    \end{pmatrix}\,.
\end{equation}
\end{widetext}
Because of the possible leakage to noncomputational levels, which we also describe by the same errors $\epsilon_{k'l', kl}$, this projected operator is generally nonunitary.
The unitarity of the operator before the projection into the computational subspace implies that
\begin{equation}
    \epsilon_{00, kl} + \epsilon_{11, kl} = \sum_{k'l' \notin \{{00}, {11}\}} \epsilon_{k'l', kl}
\end{equation}
for $kl \in \{{00}, {11}\}$
and
\begin{equation}
    \epsilon_{kl, kl} = \sum_{k'l' \ne kl} \epsilon_{k'l', kl}
    \label{eq:d3}
\end{equation}
for $kl \in \{{01}, {10}\}$, where the sum in the r.h.s. of Eq.~(\ref{eq:d3}) includes noncomputational states. Because of the probability errors, Eq.~\eqref{U_real} also contains an additional phase factor $e^{i\beta/2}$ in the off-diagonal matrix element. Because of the unitarity of the operator before the projection, the value of $\beta$ is not constant and we find that it is at least linear in probability errors, $\beta = O(\epsilon)$. 

To the linear order in the population errors $\epsilon_{k'l', kl}$, we obtain the fidelity of the real gate~\eqref{U_real}~\cite{Pedersen2007}
\begin{align}
    F  & = \frac{{\rm Tr} \left(\hat{U}_{\rm real}^\dagger \hat{U}_{\rm real}\right) + \left|\hat{U}_{\rm real}^\dagger \hat{U}_{00-11}(\pi/2, \zeta)\right|^2}{20}
    \nonumber\\
   & = 1 - \mathcal{E}_{\rm comp} - P_{\rm leak} + O\left(\epsilon^2\right)\,.
   \label{F_contributions}
\end{align}
Here
\begin{equation}\label{error-comp}
    \mathcal{E}_{\rm comp} = \frac{1}{5}\left(\epsilon^{\rm comp}_{00} + \epsilon^{\rm comp}_{01} + \epsilon^{\rm comp}_{10} + \epsilon^{\rm comp}_{11}\right)
\end{equation}
is the error due to leakage to wrong computational levels, 
where
\begin{equation}
    \epsilon^{\rm comp}_{kl} = 
    \begin{cases}
    \epsilon_{01, kl} + \epsilon_{10, kl}\quad{\rm if}\quad kl \in \{00, 11\} \\
    \epsilon_{00, kl} + \epsilon_{11, kl} + \epsilon_{lk, kl}\quad{\rm if}\quad kl \in \{01, 10\} 
    \end{cases}
\end{equation}
is the probability to leak to the wrong computational state if starting in state $\ket{kl}$. The second contribution in Eq.~\eqref{F_contributions} is the error due to the leakage to noncomputational levels, which is  given by
\begin{multline}\label{error-leakage}
    P_{\rm leak} = 1 - \frac 14{\rm Tr}\left(\hat{U}_{\rm real}^\dagger \hat{U}_{\rm real}\right) 
    \\
    =\frac{1}{4}\left(\epsilon^{\rm leak}_{00} + \epsilon^{\rm leak}_{01} + \epsilon^{\rm leak}_{10} + \epsilon^{\rm leak}_{11}\right)\,,
\end{multline}
where
\begin{equation}
    \epsilon^{\rm leak}_{kl} = 
    \begin{cases}
    \epsilon_{00, kl}  + \epsilon_{11, kl} - \epsilon_{kl}^{\rm comp} \quad{\rm if}\quad kl \in \{00, 11\} \\
    \epsilon_{kl, kl}  - \epsilon_{kl}^{\rm comp} \quad{\rm if}\quad kl \in \{01, 10\}
    \end{cases}
\end{equation}
is the probability to leak to higher states if starting in the computational state $\ket{kl}$.

Apparently, linear leakage is insufficient to explain all the errors away from optimal point. There, we need to consider another
source of error that is quadratic in $\epsilon$:
\begin{equation}\label{error-mixing}
    \mathcal{E}_{\theta} = \frac{(\epsilon_{00, 00} - \epsilon_{11, 00})^2 + (\epsilon_{00, 11} - \epsilon_{11, 11})^2}{20}
\end{equation}
This is the error in the mixture of $\ket{00}$ and $\ket{11}$ states in the Bell state.

We also calculate concurrence for the starting state $\ket{00}$ using the realistic gate operator~\eqref{U_real}. To this end, we first find the component of the final state in the computational subspace $\ket{\psi(t_{\rm gate})} =  \hat{U}_{\rm real} \ket{00}$.
We calculate $\ket{\widetilde{\psi}(t_{\rm gate})} = \hat{\sigma}_y \otimes \hat{\sigma}_y \ket{{\psi}^*(t_{\rm gate})}$
and find the concurrence~\cite{Wooters1998}
\begin{multline}\label{concurrence}
    C_{00} = \left|\left\langle\left.{\widetilde{\psi}}\right|{\psi}\right\rangle \right| 
    = 
    \left|\sqrt{(1 - 2\epsilon_{00, 00}) (1 - 2\epsilon_{11, 00})} \right.\\
    -\left. 2i\sqrt{\epsilon_{01, 00}\epsilon_{10, 00}} e^{i(\varphi_{01, 00} + \varphi_{10, 00} - \beta/2)}\right|\,.
\end{multline}
For $\epsilon_{01, 00}=0$, we have $1-C_{00}\simeq (\epsilon_{00, 00}+\epsilon_{11, 00})$.

\bibliography{literature}

\end{document}